\documentclass[aps,pra,reprint]{revtex4-2}

\usepackage{graphicx}
\usepackage{dcolumn}
\usepackage{color}
\usepackage{bm}
\usepackage{todonotes}
\usepackage{amsmath}
\usepackage{physics}
\usepackage[bookmarksopen]{hyperref}
\usepackage{mathtools}
\usepackage{comment}
\usepackage{csquotes}
\usepackage{soul}

\definecolor{darkblue}{rgb}{0.0,0.0,0.4}
\definecolor{darkgreen}{rgb}{0.0,0.4,0.0}
\definecolor{darkred}{rgb}{0.6,0.0,0.0}
\hypersetup{colorlinks,linkcolor=darkblue,citecolor=darkblue,urlcolor=darkblue}

\newcommand{\BaFOneThreeSeven}{${}^{137}$BaF}
\newcommand{\BaFOneThreeEight}{${}^{138}$BaF}

\begin{document}
\title{A platform for nuclear symmetry-violation searches\\with laser-coolable molecules carrying spinful nuclei}
\author{Tatsam Garg}
\author{Jakob Wei\ss}
\author{Tesse Tiemens}
\author{Charly Beulenkamp}
\author{Andreas Schindewolf}
\author{Tim Langen}
\email{tim.langen@tuwien.ac.at}
\affiliation{Vienna Center for Quantum Science and Technology,
Atominstitut, TU Wien, Stadionallee 2, 1020 Vienna, Austria}

\begin{abstract}
Cold heavy molecules are promising systems for exploring nuclear $\mathcal{P}$- and $\mathcal{CP}$-violating phenomena in search of new physics beyond the Standard Model. However, most proposed experimental strategies and their early realizations to date have been limited to proof-of-principle molecular species with effectively spin-zero nuclei that are not sensitive to nuclear symmetry-violating phenomena. Here, we introduce a comprehensive experimental toolbox that integrates cooling, trapping, coherent state manipulation, and a complete precision-measurement protocol that is applicable to molecules carrying relevant nuclear spins. Using ${}^{137}$Ba${}^{19}$F and nuclear-spin-dependent parity violation (NSD-PV) as representative species and benchmark application, respectively, our approach achieves a projected statistical sensitivity roughly two orders of magnitude beyond comparable molecular beams by combining techniques already demonstrated individually in current experiments. This level of precision could provide realistic experimental access not only to the enhanced NSD-PV signals arising from the heavy ${}^{137}$Ba nucleus within this molecule but also to the contributions from the lighter ${}^{19}$F nucleus, bringing direct benchmarks of nuclear \textit{ab initio} theory within reach. We further identify a candidate magic wavelength as a route to second-scale rotational coherence in future experiments. The techniques developed here can be transferred to measurements of nuclear Schiff and magnetic quadrupole moments in molecules containing deformed nuclei, establishing a general platform for laboratory searches for nuclear symmetry violations.
\end{abstract}

\maketitle

\section{Introduction}
\label{sec:introduction}

Precision tests of fundamental symmetries using molecules provide powerful probes for particle physics across a wide range of energy scales~\cite{DeMille2017,Safronova2018,DeMille2024,Hutzler2020}. At low energies, they enable studies of electroweak interactions in regimes that remain comparatively unexplored. At the same time, symmetry-violating observables can be sensitive to virtual particles and interactions at much higher energies, allowing experiments to also probe physics beyond the Standard Model (SM) at scales beyond the direct reach of current collider experiments~\cite{Hutzler2020}. 

Over the past two decades, searches for the hypothetical permanent electric dipole moment of the electron (eEDM) have provided a striking demonstration of this potential. Taking advantage of the large effective electric fields intrinsic to heavy polar molecules, these experiments have achieved some of the most stringent constraints on new $\mathcal{CP}$-violating physics to date~\cite{Hudson2011,ACME2018,Roussy2023}. 

Recent advances in molecular synthesis, cooling, and coherent control are now opening a new frontier: the use of molecules to probe symmetry-violating phenomena in the nuclear sector~\cite{ArrowsmithKron2024,Jadbabaie2026}. Such measurements promise to provide access to nuclear-spin-dependent parity violation (NSD-PV), nuclear anapole moments, Schiff moments, magnetic quadrupole moments, and other manifestations of $\mathcal{P}$- and $\mathcal{CP}$-violating physics involving nucleons~\cite{Chupp2019}. 

Accessing these observables requires molecules that contain a spin-carrying heavy nucleus and a significant electronic density at the location of this nucleus~\cite{Kozlov1995}. For molecular species with quasi-closed cycling transitions --- the enabling feature for efficient state preparation, laser cooling and quantum control --- this typically means working with fermionic, odd isotopologues, which carry more than a single nuclear spin. This gives rise to a substantially more complex hyperfine structure than is found in their bosonic even-isotopologue counterparts that have been used in most molecular precision-measurement, cooling and trapping experiments to date~\cite{Fitch2021,Kogel2021,Kogel2025lasercooling,Zeng2023}. As a result, the experimentally most relevant species for nuclear symmetry-violation studies remain largely unexplored in terms of state-of-the-art state-preparation, cooling and control techniques.

In this work, we develop and analyze a complete experimental toolbox for precision measurements with such laser-coolable spinful molecules. Using \BaFOneThreeSeven\ as representative species, we employ NSD-PV as a benchmark application to illustrate the potential of this approach. More broadly, this outlines a general pathway toward precision searches for nuclear symmetry violations and other manifestations of physics beyond the SM.

\section{Nuclear-spin-dependent parity violation as a benchmark application}
\label{sec:nsdpv}

Among the broad range of precision measurements enabled by spinful molecules, nuclear-spin-dependent parity violation (NSD-PV) provides a particularly attractive benchmark. The measurement protocol is well established~\cite{Flambaum1985,Kozlov1995,DeMille2008,Norrgard2019,Karthein2024}, quantitative predictions exist for several candidate species~\cite{DeMille2008,Borschevsky2013,Hao2018,Hao2020,Isaev2010}, and proof-of-principle experiments using molecular beams have demonstrated the underlying concepts~\cite{Altuntas2018}. 

NSD-PV probes parity-violating weak electron--nucleon and nucleon--nucleon interactions through contributions dominated by vector-electron--axial-nucleon ($V_eA_n$) weak neutral-current couplings and the nuclear anapole moment~\cite{Flambaum1997,Haxton2001,Haxton2002}. Precision measurements of NSD-PV promise stringent benchmarks for nuclear \textit{ab initio} theory~\cite{Borschevsky2013,Hao2018,Hao2020}, enable the extraction of poorly known weak coupling constants~\cite{DeMille2008,Haxton2002}, may constrain light bosons from the dark sector~\cite{Stadnik2014,Gaul2026}, and offer valuable input for resolving longstanding discrepancies in our understanding of hadronic parity violation~\cite{Gardner2026}. They therefore provide a unique connection between atomic, molecular, particle, and nuclear physics. 

The relevant observables arise from weak-interaction-induced mixing between two states of opposite parity, which scales inversely with the energy separation of these states. Building on landmark atomic measurements~\cite{Bouchiat1997,Wood1997,Safronova2018,Antypas2019} and subsequent proof-of-principle molecular-beam experiments~\cite{Cahn2014,Altuntas2018}, measurement schemes maximizing this mixing have been established. In particular, in diatomic molecules, the rotational structure naturally produces pairs of opposite-parity states with very small energy separations, which can further be tuned to near degeneracy using external magnetic fields, enhancing signals by many orders of magnitude~\cite{Kozlov1995,DeMille2008}. Nevertheless, the resulting parity mixing remains small, placing stringent requirements on the identification of suitable state pairs, their coherence times, and the control and probing of the molecules inside the applied electromagnetic fields. As in atomic clocks and other precision spectroscopy experiments, laser cooling and closely related optical control techniques provide a natural route to address these requirements in molecules~\cite{Ludlow2015,Safronova2018}.

\begin{figure}[tb]
    \centering
    \includegraphics[width=1\linewidth]{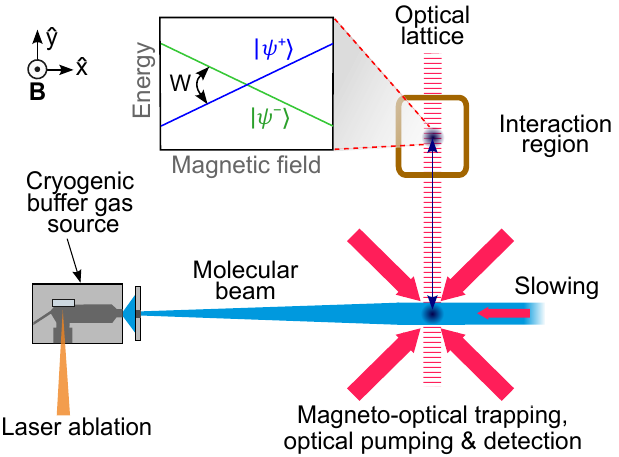}
    \caption{Schematic of the proposed experimental apparatus to measure nuclear-spin-dependent parity violation. The apparatus implements the toolbox developed in Secs.~\ref{sec:measurement-principle}--\ref{sec:dipole-trapping-and-transport}. Molecules are created by laser ablation in a cryogenic buffer-gas source. The resulting molecular beam is slowed and subsequently captured in a magneto-optical trap (MOT), with pink arrows indicating the corresponding laser beams. From there, the molecules are optically pumped into a suitable science state, loaded into a movable optical lattice, and transported into the interaction region. Here, a stable magnet Zeeman-tunes the opposite-parity science states $\ket{\psi^\pm}$ to near degeneracy with a magnetic field $\mathbf{B}$, enhancing their mixing by a weak matrix element $W$ by many orders of magnitude~\cite{Kozlov1995,DeMille2008,Altuntas2018}. The mixing is detected via optical cycling after the molecules have been transported back to the MOT region. The science goal is to extract $W$, which encodes fundamental parity-violating effects of the weak interaction.}
    \label{fig:setup}
\end{figure}

An example apparatus that exploits this strategy is introduced in Fig.~\ref{fig:setup}. Molecules are cooled and trapped, optically pumped, and transported into a science region, where precision measurements are performed. We focus our analysis on \BaFOneThreeSeven, whose spin-carrying 
$^{137}$Ba nucleus combines a comparatively large predicted NSD-PV signal~\cite{Hao2018} with excellent prospects for laser cooling and trapping~\cite{Kogel2021,Kogel2025lasercooling,Kogel2026}.

In the following sections, we identify the corresponding pairs of opposite-parity science states in \BaFOneThreeSeven\ and develop the full experimental toolbox to perform the precision measurement. Although the level-crossing scheme itself is specific to NSD-PV, the capabilities it demands --- large trapped-molecule numbers, high-fidelity state preparation, long coherence times, and control of systematic shifts --- are common to nuclear precision measurements using molecules in general~\cite{ArrowsmithKron2024,Jadbabaie2026,Ho2023}. \BaFOneThreeSeven\ is therefore representative of the broader class of spinful molecules relevant to future searches for nuclear symmetry violation, including molecules containing nuclei with static octupole deformation, such as neutron-rich barium isotopes around $^{145}$Ba~\cite{Flambaum2014,Bucher2016} or the odd isotopes $^{223}$Ra and $^{225}$Ra of radium~\cite{Udrescu2021}, as well as quadrupole-deformed nuclei such as $^{173}$Yb~\cite{Ho2023}.

\section{Measurement principle and identification of science states}
\label{sec:measurement-principle}

We begin by examining the rotational, hyperfine, and Zeeman structure of \BaFOneThreeSeven\ and identifying suitable science state-pairs for NSD-PV measurements.

In the absence of external fields, the spin and rotational structure of a single vibrational level in the $^2{\Sigma^+}$ electronic ground state of the odd isotopologues of BaF is described by the effective Hamiltonian
\begin{equation}
\label{eqn:molecule-hamiltonian}
    \begin{split}
        H_{{}^2{\Sigma^+}} &= B\mathbf{N}^2 - D\mathbf{N}^4 + \gamma \mathbf{N} \cdot \mathbf{S} \\
        &\quad + b_{\text{Ba}}\,\mathbf{I}_{\text{Ba}} \cdot \mathbf{S} \\
        &\quad + c_{\text{Ba}}\left[\left(\mathbf{I}_{\text{Ba}}\cdot \mathbf{\hat{n}}\right)\left(\mathbf{S}\cdot \mathbf{\hat{n}}\right) - \frac{1}{3}\mathbf{I}_{\text{Ba}} \cdot \mathbf{S}\right] \\
        &\quad + eq_0Q_{\text{Ba}} \frac{3\left(\mathbf{I}_{\text{Ba}}\cdot \mathbf{\hat{n}}\right)^2 - \mathbf{I}_{\text{Ba}} \cdot \mathbf{I}_{\text{Ba}}}{4I_{\text{Ba}}(2I_{\text{Ba}} - 1)} \\
        &\quad + b_{\text{F}}\,\mathbf{I}_{\text{F}} \cdot \mathbf{S} \\
        &\quad + c_{\text{F}}\left[\left(\mathbf{I}_{\text{F}}\cdot \mathbf{\hat{n}}\right)\left(\mathbf{S}\cdot \mathbf{\hat{n}}\right) - \frac{1}{3}\mathbf{I}_{\text{F}} \cdot \mathbf{S}\right],
    \end{split}
\end{equation}
where $\mathbf{N}$, $\mathbf{S}$, and $\mathbf{I}$ are the rotational, electronic spin, and nuclear spin angular momenta~\cite{Kogel2024fermispectroscopy}, with corresponding quantum numbers $N$, $S$, and $I$, and $\mathbf{\hat{n}}$ is the unit vector along the molecule's internuclear axis. The subscripts label the barium (Ba) and fluorine (F) nuclei. Moreover, $B$ and $D$ are the rotational and centrifugal distortion correction constants, $\gamma$ is the spin-rotation coupling constant, and $b$ and $c$ characterize the hyperfine couplings. The electric-field gradient of the valence electron, $eq_0$, and the nuclear quadrupole moment of the Ba nucleus, $Q_{\text{Ba}}$, contribute to the electric quadrupole term. These parameters have previously been precisely measured with optical and microwave spectroscopy, while additional terms that describe couplings of nuclear spin and rotational degrees of freedom have been determined to be negligible~\cite{Kogel2024fermispectroscopy, Ryzlewicz1982,Preston2026}.

The rotational constant $B$ is much larger than all other couplings in Eq.~\ref{eqn:molecule-hamiltonian} and $N$ is a good quantum number for labeling the eigenstates with energies $E_N = BN(N+1)$ and well-defined parity $P=(-1)^N$. The $N=0$ and $N=1$ rotational manifolds therefore exhibit opposite parity. 
\begin{figure}
    \centering
    \includegraphics[width=1.0\linewidth]{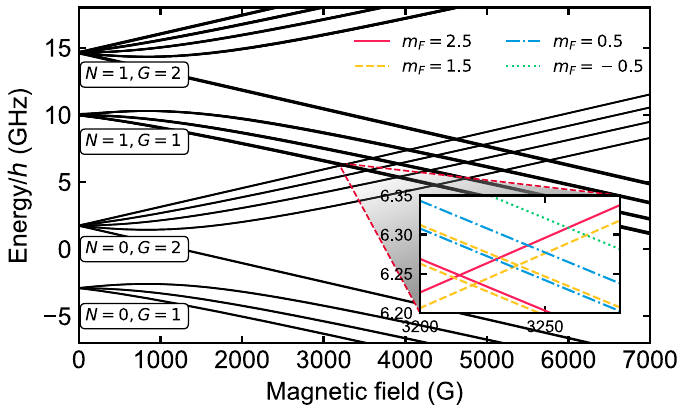}
    \caption{Structure of \BaFOneThreeSeven\ in a magnetic field. Levels belonging to the $N=0$ and $N=1$ rotational states undergo crossings between $3000$ and $6000\,$G. A total of $192$ individual crossings occur, which are organized in $16$ clusters. As shown in the inset, each cluster contains $12$ individual crossings, each associated with individual hyperfine levels $m_F$. The representative science crossing examined in detail in this work is between the $m_F=5/2$ levels at approximately $3215\,$G, denoted by the solid pink line in the inset. The remaining state labels are introduced in the main text.}
    \label{fig:Zeeman-crossings}
\end{figure}
For the NSD-PV measurement they would be tuned to near-degeneracy using an external magnetic field $\mathcal{\mathbf{B}}=\mathcal{B}\mathbf{\hat{z}}$, where $\mathbf{\hat{z}}$ is a unit vector in the laboratory frame. The Zeeman shifts are governed by the effective Zeeman Hamiltonian
\begin{equation}
\label{eqn:Zeeman-Hamiltonian}
    \begin{split}
        H_Z &= g_S \mu_B \mathcal{B} \textrm{T}^1_{p=0}(\mathbf{S}) + g_l\mu_B \mathcal{B}\sum_{q=\pm 1}\mathcal{D}^{(1)}_{p=0,q}(\omega)^*\textrm{T}^1_{q}(\mathbf{S})\\
        &\quad -g_r \mu_B \mathcal{B} \textrm{T}^1_{p=0}(\mathbf{N}) - g_{N_{\text{F}}} \mu_N \mathcal{B}\textrm{T}^1_{p=0}\left(\mathbf{I}_{\text{F}}\right) \\
        &\quad -g_{N_{\text{Ba}}} \mu_N \mathcal{B}\textrm{T}^1_{p=0}\left(\mathbf{I}_{\text{Ba}}\right) 
    \end{split}
\end{equation}
adapted from Ref.~\cite{Brown2003} for a ${}^2\Sigma$ state with two nuclear spins. Here, the subscript $q$ denotes the molecule-frame components, $p=0$ denotes the component along the laboratory-frame magnetic field, and $\mathcal{D}^{(1)}_{p=0,q}(\omega)$ is the Wigner D-matrix that rotates operator components between these two frames. For the g-factors $g_S$, $g_l$, and $g_r$ we use recently measured values for \BaFOneThreeEight\ \cite{Cahn2014} as approximations to study the structure of level crossings in \BaFOneThreeSeven. Figure~\ref{fig:Zeeman-crossings} illustrates the resulting Zeeman-tuned crossings between the $N=0$ and $N=1$ manifolds.

The preparation of the molecules and the actual precision measurements benefit from very different magnetic-field regimes. It is therefore useful to distinguish the quantum numbers that describe the molecular eigenstates in the low- and high-field limits.

At low magnetic fields, where laser slowing, cooling, trapping, and optical state preparation will be performed, the energy eigenstates are well described in the coupled Hund's case $(b_{\beta S})$ 
basis 
\begin{equation*}
\ket{N,G,F_1,F,m_F}, 
\end{equation*}
where $\textbf{G}=\textbf{I}_{\text{Ba}}+\textbf{S}$, $\textbf{F}_1=\textbf{N}+\textbf{G}$, and $\textbf{F}=\textbf{F}_1+\textbf{I}_{\text{F}}$ are the two corresponding intermediate angular momenta and the total angular momentum, respectively. Here and in the following, $m$ quantum numbers denote the eigenvalues of the laboratory-frame projections of these angular momenta along the quantization axis set by the magnetic field. 

As the magnetic field increases, the Zeeman shifts dominate over spin-rotation and hyperfine interactions, decoupling the angular momenta. Near the level crossings, the eigenstates are thus approximately described in the fully decoupled basis 
\begin{equation*}
\ket{N,m_N}\ket{S,m_S}\ket{I_{\text{Ba}},m_{I_{\text{Ba}}}}\ket{I_{\text{F}},m_{I_{\text{F}}}}.
\end{equation*}
Throughout the entire field range, the laboratory-frame projection of the total angular momentum $
    m_F = m_N + m_S + m_{I_{\text{Ba}}} + m_{I_{\text{F}}}$
remains a good quantum number. 

Near the crossings, the electronic Zeeman interaction dominates, such that the upward- and downward-sloping manifolds belong predominantly to $m_S=+1/2$ and $m_S=-1/2$ states, respectively. 

In total, we find $192$ individual level crossings, which are shown in Fig.~\ref{fig:Zeeman-crossings}. The $4\times4=16$ clusters of crossings spread between $3000$\,G to $6000$\,G are produced by the $2I_{\text{Ba}}+1=4$ degrees of freedom arising from $I_{\text{Ba}}=3/2$ in each of the two rotational manifolds. Within each cluster, as illustrated by the inset in Fig.~\ref{fig:Zeeman-crossings}, we have $(2N+1)\times(2I_{\text{F}}+1)$ states from each rotational manifold, yielding $2\times6=12$ crossings for $I_{\text{F}}=1/2$, $N=0,1$. 

Among this multitude, a pair of states that simultaneously maximizes the NSD-PV matrix element and enables efficient state preparation needs to be identified. The latter can often be implemented most straightforwardly for stretched states, while the former is determined by the matrix elements of the effective NSD-PV Hamiltonian~\cite{Flambaum1985} 
\begin{equation}
\label{eqn:NSD-PV-Hamiltonian}
    H^{\text{eff}}_P = \kappa \hbar W_P (\mathbf{S}\times\mathbf{\hat{n}}) \cdot \mathbf{I}/{I}
\end{equation}
summed over all resolved nuclear spins. The operator $(\mathbf{S}\times\mathbf{\hat{n}})\cdot\mathbf{I}/I$ is the generic pseudoscalar that can be constructed from the electron spin, the
internuclear axis, and the nuclear spin, so that any specific NSD-PV mechanism enters only through the coefficients $\kappa$ and $W_P$. Here, $W_P$ is a measure of the valence electron's overlap with the nuclei and is supplied by \textit{ab initio} theory or semi-empirical calculations, and $\kappa$ encodes the various NSD-PV effects discussed above.

The NSD-PV measurement between states of opposite parity, $\ket{\psi^-}$ and $\ket{\psi^+}$, probes $i\hbar W = \bra{\psi^-} H^{\text{eff}}_P \ket{\psi^+} = \kappa \hbar W_P \tilde C$, where $\tilde C = \bra{\psi^-} (\mathbf{S}\times\mathbf{\hat{n}}) \cdot \mathbf{I}/I\ket{\psi^+}$. $H^{\text{eff}}_P$ is block-diagonal in $m_F$, which means that only the matrix elements between states with $m_F = m_{F}'$ are non-zero. We identify all level crossings that satisfy this condition and find the associated science states. The corresponding dominant fully decoupled basis labels, matrix elements $\tilde{C}$ and crossing positions are tabulated in Appendix~\ref{app:level-crossings-table}. The most promising crossing is found at $\mathcal{B} \approx 3215$\,G between
\begin{equation*}
    \ket{\psi^+}=\ket{0,0}\ket{1/2, +1/2}\ket{3/2, +3/2}\ket{1/2,+1/2}
\end{equation*}
and
\begin{equation*}
    \ket{\psi^-}=\ket{1, +1}\ket{1/2,-1/2}\ket{3/2,+3/2} \ket{1/2,+1/2}
\end{equation*}
with $\tilde C \approx -0.44\,i$. Adiabatically tracking these states to the low-field limit, they correspond to the Hund's case $(b_{\beta S})$ stretched states 
\begin{equation*}
\ket{\psi_\mathrm{\mathcal{B}=0}^+}=\ket{N=0,G=2,F_1=2,F=5/2,m_F=+5/2},
\end{equation*}
and 
\begin{equation*}
\ket{\psi_\mathrm{\mathcal{B}=0}^-}=\ket{N=1,G=1,F_1=2,F=5/2,m_F=+5/2}.
\end{equation*}

The negative-parity state $\ket{\psi_\mathrm{\mathcal{B}=0}^-}$ exhibits significant potential for efficient optical pumping and state preparation since it is a stretched state of the $G=1$ manifold involved in a rotationally-closed optical cycling transition~\cite{Kogel2025lasercooling}. A detailed scheme for optical pumping into this state that is based on this observation is developed in Sec.~\ref{sec:state-preparation}.

Other crossings, e.g.\ for a comprehensive exploration of systematics, can be found along similar lines. In particular, we emphasize that the procedure outlined here for identifying suitable science states is general -- only the effective Hamiltonian describing the target 
observable, e.g.\ Eq.~\ref{eqn:NSD-PV-Hamiltonian} for NSD-PV, changes between science goals and species, while the procedure itself is common to a broad class of molecules. For example, radioactive $^{223}$RaF, which has been created and studied in a series of recent experiments~\cite{GarciaRuiz2020,Udrescu2021}, shares the same nuclear spins ($I_{\mathrm{Ra}} = 3/2$, $I_{\mathrm{F}} = 1/2$) and is even described by the same Hamiltonian of Eq.~\ref{eqn:molecule-hamiltonian}, so its state structure and science-state selection for NSD-PV experiments carry over directly, up to a rescaling of the molecular constants.

The general measurement scheme for $i\hbar W$ follows the Stark-interferometry method described in~\cite{DeMille2008,Altuntas2018,Norrgard2019}. As discussed above, a static magnetic field $\mathbf{B}$ is used to tune the states of opposite parity, $\ket{\psi^+}$ and $\ket{\psi^-}$, to near-degeneracy with their residual splitting denoted by $\Delta$, with units of angular frequency. The wavefunction of the two-level system is then described by
\begin{equation}
    \ket{\psi(t)}= e^{-i \Delta t}c_- (t)\ket{\psi^-}+c_+ (t)\ket{\psi^+},
\end{equation}
where we initialize $c_-(0)=1$ and $c_+(0)=0$ using parity-selective depletion. A uniform oscillating electric field $\mathbf{E}=\mathcal{E}\mathrm{cos}(\omega t)\mathbf{\hat{z}}$ that is oriented along the magnetic-field direction induces dipole transitions between the science states due to their admixed nature~\cite{DeMille2008}. The time evolution of $\ket{\psi(t)}$ is calculated in first-order time-dependent perturbation theory using the effective Hamiltonian
\begin{equation}
\label{eqn:Stark-interferometry-hamiltonian}
    H_{\pm} = \begin{pmatrix}
        \hbar\Delta & d \mathcal{E}\mathrm{cos}(\omega t) + i \hbar W \\
        d \mathcal{E}\mathrm{cos}(\omega t) - i \hbar W & - \alpha'\mathcal{E}^2\mathrm{cos}^2(\omega t)/2
    \end{pmatrix},
\end{equation}
where $d$ is the transition dipole moment. The term with the differential polarizability $\alpha'$ only accounts for the time-dependent contribution from the Stark-mixing electric field, and has been shown to have a negligible contribution for the weak electric-field strengths $\mathcal{E}$ typically employed in such measurements~\cite{Cahn2014}. Thus, in the time-dependent Hamiltonian, we assume $\alpha'=0$. 

In the measurement, the weak interaction mixes population from $\ket{\psi^-}$ into the initially empty state $\ket{\psi^+}$, which is captured by the signal
\begin{equation}
\label{eqn:observable}
    S = |c_+ (t)|^2 \approx 4 \left[ 2 \frac{W}{\Delta} \frac{d \mathcal{E}}{\hbar \omega} + \left( \frac{d \mathcal{E}}{\hbar \omega} \right)^2 \right] \mathrm{sin}^2\left( \frac{\Delta t}{2} \right)
\end{equation}
in the limit where $\hbar W \ll d \mathcal{E}$ and $\Delta \ll \omega$. The first term in the square brackets is odd under the reversal of $\mathbf{E}$, whereas the second term is even, which produces the asymmetry
\begin{equation}
\label{eqn:asymmetry}
    \mathcal{A} =\frac{S(+\mathcal{E})-S(-\mathcal{E})}{S(+\mathcal{E})+S(-\mathcal{E})} \approx 2 \frac{W}{\Delta} \frac{\hbar\omega}{d\mathcal{E}}.
\end{equation}
This expression is directly proportional to the desired matrix element $W$, and enhanced by the inverse dependence on $\Delta$, which can be made very small through the Zeeman tuning. This results in an enhancement of the NSD-PV effects by many orders of magnitude in a large number of suitable molecular species~\cite{DeMille2008}. 

Importantly, we observe that a measurement for $W$ is only as precise as the control over $\Delta$, $\mathcal{E}$, and $\omega$. The electric-field amplitude $\mathcal{E}$ and the coupling frequency $\omega$ can be controlled with high precision using commercial function generators that easily provide amplitude fluctuations $
\delta\mathcal{E}<1\,$mV/cm and sub-Hz-level frequency stability for sine waves in the $10$--$100$\,kHz range. 

This leaves the detuning $\Delta$ as the dominant experimental source of uncertainty. Consequently, the achievable sensitivity is ultimately limited by drifts of $\Delta$, as well as by its temporal fluctuations and spatial variations, which we denote by $\delta\Delta$. This places stringent requirements on the control of magnetic and any other fields that may influence these parameters throughout the measurement. 

We note in particular that if the molecules are held in a trap for the measurement, differential shifts between $\ket{\psi^\pm}$ can lead to significant contributions to both uncertainty and fluctuations in $\Delta$. Nevertheless, trapping of molecules offers two key advantages: First, it enables substantially improved control of the electromagnetic environment compared to a molecular beam. Second, it extends the available interrogation time by several orders of magnitude. A detailed analysis of these aspects is presented in Sec.~\ref{sec:dipole-trapping-and-transport}. 

Realizing these advantages requires a complete experimental platform capable of producing, trapping, preparing, and coherently interrogating the molecular ensemble. The remainder of this work develops this platform, following the experimental sequence from molecule production and slowing to laser cooling, state preparation, and precision measurement.

\section{Laser cooling and trapping}
\label{sec:laser-cooling-and-trapping}

The first stages of the experimental sequence shown in Fig.~\ref{fig:setup} consist of molecule production, slowing, and trapping. Molecules suitable for laser cooling and precision-measurement experiments are typically produced using cryogenic buffer-gas sources, which generate molecular beams with forward velocities on the order of $100$--$200$\,m/s~\cite{Hutzler2012}, well above the capture velocities of magneto-optical traps~\cite{Fitch2021}. Bringing these molecules to rest therefore requires optical slowing followed by efficient laser cooling and trapping. 

Both laser slowing and cooling in molecules are based on realizing a quasi-closed optical cycle, which for BaF molecules can be achieved using the $X(\nu=0,N=1) \rightarrow A(\nu'=0,J'=1/2^+)$ electronic transition~\cite{Fitch2021,Albrecht2020,Kogel2025lasercooling}. Here, $X$ denotes the electronic ground state described by Eq.~\ref{eqn:molecule-hamiltonian} and $\nu$ is the vibrational quantum number. The electronically excited state $A$ is described by Hund's case $(a)$, where the total angular momentum without nuclear spins is denoted by $J$ and parity by the exponent $+$. 

The choice of rotational manifolds in this transition suppresses rotational leakage through parity selection rules, since the lowest positive-parity excited-state manifold $J'=1/2^+$ can decay only to negative-parity ground states, 
which combined with angular-momentum selection rules leaves only $N=1$. Residual vibrational leakage can be mitigated using additional repump lasers that address similar transitions with $\nu \rightarrow \nu'=\nu-1$ and $\nu\geq1$~\cite{Fitch2021}. This scheme is generic and applies to both even and odd isotopologues of BaF, but must additionally accommodate the different hyperfine complexity of the two, using multiple laser frequency components generated by electro- and acousto-optic modulation~\cite{Kogel2025serrodynes}.

Based on these principles, transverse laser cooling of a \BaFOneThreeSeven\ beam has been demonstrated~\cite{Kogel2025lasercooling}. However, efficient slowing and magneto-optical trapping of spinful odd isotopologues --- required to capture the molecules from such a beam --- have yet to be realized~\cite{Langen2023,DeMille2024,Fitch2021}.

The main obstacle is the reduced optical scattering rate expected due to the additional hyperfine structure in odd isotopologues, which, combined with the low recoil velocity of heavy species, renders conventional slowing based on the scattering force highly inefficient. Additionally, this mechanism suffers from transverse beam pluming due to spontaneous-emission recoil, reducing the loading efficiency of subsequent cooling stages. These challenges are not specific to \BaFOneThreeSeven\ but generic to heavy spinful molecules. In the following, we show that they can be overcome using bichromatic-force slowing, which relies on stimulated rather than spontaneous photon scattering, combined with conveyor-belt magneto-optical trapping. Both of these techniques can be implemented with efficiencies comparable to those achieved for even isotopologues~\cite{Zeng2024,Zeng2026,Kogel2026}.

\begin{figure}
    \centering
        \includegraphics[width=1\linewidth]{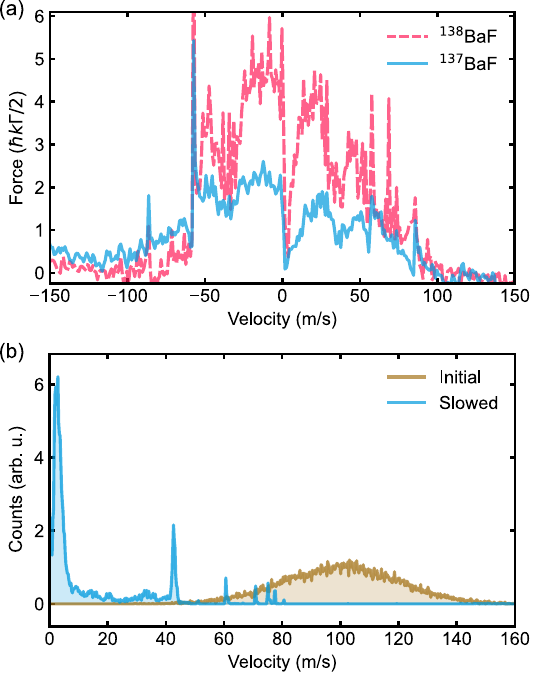}
    \caption{Bichromatic forces as a tool for efficient slowing and velocity compression. (a) Simulated bichromatic forces in \BaFOneThreeSeven\ (blue solid line) and \BaFOneThreeEight\ (pink dashed line) as a function of particle velocity for a bichromatic detuning of $\delta_{\text{BC}}=\pm 2\pi \times 200$\,MHz, laser intensity $I_{\text{BC}}\approx26$\,W/cm$^2$ per frequency component per direction, and bichromatic phase $\phi_{\text{BC}}=\pi/2$. While the overall force magnitude is reduced in the odd isotopologue \BaFOneThreeSeven, it still significantly exceeds the maximum scattering force $\hbar k\Gamma/2$ achievable in an equivalent idealized two-level system with linewidth $\Gamma$ and wavelength $\lambda=2\pi/k$ over a wide range of velocities. In particular, it is orders of magnitude higher than scattering forces taking into account the full hyperfine structure~\cite{Kogel2025lasercooling,Kogel2026}. The force profiles of both isotopologues naturally produce strong velocity compression during slowing owing to a hyperfine-structure-induced dip in the force magnitude near zero velocity. (b)~Monte Carlo simulation of chirped and phase-compensated bichromatic-force slowing of \BaFOneThreeSeven\ for a slowing time of 3\,ms, slowing distance of 15\,cm, and frequency chirp over 240\,MHz. The initial velocity distribution centered at 100\,m/s (brown area) is efficiently slowed to velocities under 5\,m/s (blue area), well within the capture velocity of a conveyor-belt MOT. The narrow peak in velocity demonstrates the efficient velocity compression in a moving frame resulting from the dip in force magnitude illustrated in (a).}
    \label{fig:BCF}
\end{figure}

\begin{figure*}
    \centering
        \includegraphics[width=1\linewidth]{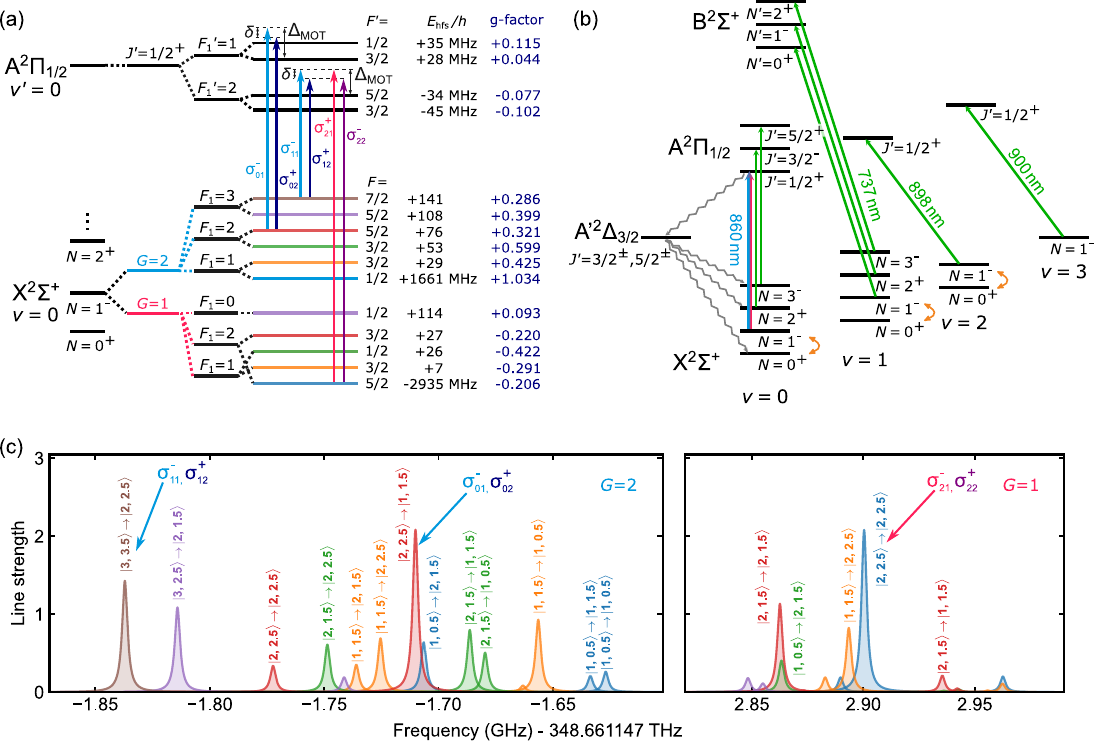}

    \caption{Triple conveyor-belt MOT scheme for cooling and trapping of \BaFOneThreeSeven. (a) Three pairs of oppositely circular polarized frequency components (handedness of polarization denoted by the superscript) address three dominant transitions with a global blue detuning $\Delta_{\text{MOT}}$ and two-photon detuning $\delta$ between frequency components within each pair. For a total laser intensity, $I_{\text{tot}}$, the intensity is split between $[\sigma_{01}, \sigma_{02}, \sigma_{11}, \sigma_{12}, \sigma_{21}, \sigma_{22}]$ components as $[0.2, 0.2, 0.15, 0.15, 0.15, 0.15]\times I_{\text{tot}}$. The hyperfine energy shifts $E_{\text{hfs}}$ relative to the respective rotational levels and the effective g-factors for the linear-Zeeman-effect regime are given for each hyperfine level $F$. (b)~The realization of this optical cycle in \BaFOneThreeSeven\ involves repumping to minimize vibrational leakage from imperfect Franck-Condon factors (FCFs), as well as rotational leakage from decays via the metastable $A'{}^2\Delta$ state and hyperfine-induced mixing processes in the excited states. The superscript to the rotational labels indicates parity and straight arrows are laser-driven transitions with the given wavelengths. For a MOT, vibrational repumping for up to $\nu=3$, and rotational repumping up to $N=3$ in $\nu=0$ and $\nu=1$ is required to sufficiently close the optical cycle. States with $N \ge 1$ can be optically repumped, while microwaves (indicated by bent orange arrows) can be used to remix populations between $N=0$ and $N=1$. States which do not participate in this repumping scheme have been omitted for clarity. (c) Hyperfine spectra of the two $G$-manifolds of the cooling transition with the key transition frequencies marked by arrows, following the labels for frequency components in (a). Colors indicate the participating hyperfine level in the electronic ground state according to (a), and transition labels give the associated hyperfine quantum numbers $\ket{F_1, F}\rightarrow \ket{F_1', F'}$ for the ground and excited states. During optical pumping and laser cooling, various specifically synthesized optical spectra are used to address the  molecular transitions~\cite{Kogel2025serrodynes,Holland2021}.}
    \label{fig:MOT-scheme}
\end{figure*}

\subsection{Bichromatic-force slowing}

 The limitations of scattering-force slowing can be circumvented by using the stimulated-emission-based bichromatic force (BCF), where the interference of pairs of beams with opposite direction and pairwise detuning $\delta_{\text{BC}}$ leads to efficient coherent exchange of photons~\cite{Partlow2004,Aldridge2016}. The resulting forces are no longer limited by the scattering rates of the molecules but can, in principle, become much larger. Their advantages have been demonstrated experimentally for the slowing of atoms~\cite{Chieda2012, Soeding1997} and the deflection of molecular beams~\cite{KozyryevBichromatic2018,GalicaDeflection2018}.

Fig.~\ref{fig:BCF}a compares the simulated BCF for \BaFOneThreeSeven\ and \BaFOneThreeEight\ calculated using optical Bloch equations (OBEs) with the framework developed in Ref.~\cite{Kogel2026}. The force magnitude exceeds the maximum scattering force achievable in an equivalent two-level system over a wide range of velocities for both isotopologues. More importantly, it is orders of magnitude larger than the realistic scattering forces obtained when the full hyperfine structure of odd isotopologues is taken into account~\cite{Kogel2025lasercooling,Kogel2026}.

In our scheme for \BaFOneThreeSeven, the bichromatic light is centered at the hyperfine transition $X^2\Sigma^+(\nu=0, N=1,G=2,F_1=2,F=5/2) \rightarrow A^2\Pi_{1/2}(\nu'=0,J'=1/2^+,F_1'=1,F'=3/2)$ within the $G=2$ manifold. The bichromatic components are detuned by $\delta_{\text{BC}}= \pm 2\pi \times 200$\,MHz, which fully covers the hyperfine structure caused by the fluorine nuclear spin. 

The additional $G=1$ manifold of the $X^2\Sigma^+(\nu=0, N=1)$ state, which arises from the interactions between the electron spin and the barium nuclear spin, requires weak off-resonant repumping to mitigate any dark states in which the molecules may end up after undesired spontaneous emission. Thus, we include a weak frequency component with an intensity of 0.3\,W/cm$^2$ near the $X^2\Sigma^+(\nu=0, N=1,G=1,F_1=2,F=3/2) \rightarrow A^2\Pi_{1/2}(\nu'=0,J'=1/2^+,F_1'=2,F'=5/2)$ transition. 

For efficient slowing, the bichromatic light can be chirped to decelerate the rest frame of the counter-propagating beams and the bichromatic phase can be compensated for the longitudinal motion of the molecules along the beams~\cite{Chieda2012}. Figure~\ref{fig:BCF}b illustrates a Monte Carlo simulation of this scheme in \BaFOneThreeSeven\ with a slowing time and distance of 3\,ms and 15\,cm respectively for an initial velocity distribution centered around 100\,m/s. 

The molecules are efficiently slowed to near standstill while getting compressed into a narrow velocity distribution. This velocity compression is a signature of cooling in a moving frame, produced by the hyperfine-structure-induced dip in the force profile illustrated in Fig.~\ref{fig:BCF}a near the rest-frame zero velocity. As a result, \BaFOneThreeSeven\ molecules can be brought to MOT capture velocities within tens of centimeters of slowing distance, substantially improving the loading efficiency of subsequent trapping stages.

Taken together, these simulations demonstrate that bichromatic-force slowing remains highly efficient in \BaFOneThreeSeven\ despite its substantially increased hyperfine complexity, delivering molecules into the capture range of the conveyor-belt MOT developed below. Since the bichromatic force inherently bypasses the hyperfine structure challenge by covering all relevant states in parallel, the approach applies directly to other heavy species such as RaF and YbF, where conventional radiative slowing is equally challenging~\cite{ArrowsmithKron2024,Athanasakis2025}.

\begin{figure*}[th]
    \centering
        \includegraphics[width=\linewidth]{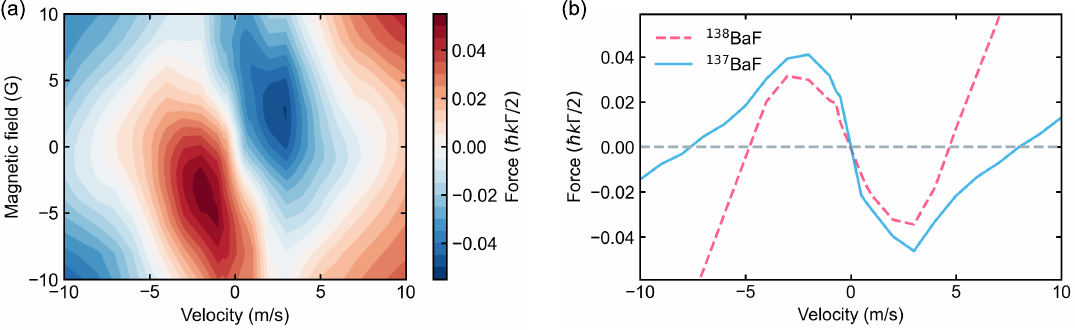}
        
    \caption{Results of the MOT simulations using the triple conveyor-belt scheme proposed in Fig.~\ref{fig:MOT-scheme}. (a) We show the simulated radial force over varying on-axis particle velocity and magnetic-field strength in the three-dimensional conveyor-belt MOT of \BaFOneThreeSeven. The simulation uses $I_{\text{tot}}=0.4$\,W/cm$^2$ per laser beam direction, single-photon detuning $\Delta_{\text{MOT}} = + 2\pi \times 10$\,MHz, and two-photon detuning $\delta=2 \pi \times 3$\,MHz. (b) Slice of the cooling force at the MOT center. Results for \BaFOneThreeSeven\ (blue solid line) are compared with those for \BaFOneThreeEight\ (pink dashed line) simulated with the experimental parameters from the directly loaded MOT experiment in Ref.~\cite{Zeng2026}, with $I_{\text{tot}}=0.2$\,W/cm$^2$ per laser beam direction, in ratio with the number of frequency components needed for this species. The results suggest comparable forces between the two isotopologues with a similar capture velocity of $\sim 5$\,m/s inferred from classical-trajectory simulations at these intensities, indicating favorable conditions for direct loading of \BaFOneThreeSeven\ molecules into a conveyor-belt MOT with the proposed scheme.}
    \label{fig:MOT-results}
\end{figure*}

\subsection{Magneto-optical trapping}

Similarly to laser slowing, the hyperfine structure of \BaFOneThreeSeven\ also substantially reduces the effectiveness of conventional Doppler forces used for magneto-optical trapping. This limitation motivates the use of a blue-detuned conveyor-belt MOT based on sub-Doppler forces, following the recent successful demonstration for \BaFOneThreeEight~\cite{Zeng2024,Zeng2026}. Unlike conventional red-detuned MOTs, the restoring force in this scheme arises predominantly from magnetically induced dark states, making it considerably more robust against the reduced scattering rates encountered in spinful molecules. These dark states arise from the interference of suitably polarized and detuned pairs of laser beams~\cite{Yu2026,Lyu2026}.

Extending this approach to the odd isotopologue is nevertheless non-trivial. The additional nuclear spin increases the number of hyperfine manifolds that contribute to the cooling dynamics, requiring the trapping forces from multiple manifolds to be balanced while maintaining a practical optical setup. 

Building on our previous work on transverse cooling~\cite{Kogel2025lasercooling,Kogel2026,Kogel2025serrodynes}, we identify the subset of hyperfine transitions that dominate the cooling dynamics and condense the optical cycling scheme to the corresponding transitions. This concentrates the available laser power onto the most relevant transitions, suppresses competition between different force contributions, and yields a tractable experimental implementation.

We solve OBEs for this system in three spatial dimensions with the configuration illustrated in Fig.~\ref{fig:MOT-scheme}a. We identify three hyperfine transitions connected to the states $\ket{G=2,F_1=2,F=5/2}$, $\ket{G=2,F_1=3,F=7/2}$, and $\ket{G=1,F_1=2,F=5/2}$ in the $X^2\Sigma^+(\nu=0,N=1)$ manifold that must each be addressed by an independent conveyor-belt frequency pair to generate sufficient trapping forces. Replacing any one of these components by a simple repumping frequency substantially reduces the simulated restoring force, requiring considerably higher laser powers to achieve comparable capture velocities. This observation is consistent with our experimental investigations of optical cycling in \BaFOneThreeSeven~\cite{Kogel2025lasercooling}. The corresponding transition frequencies are indicated in the hyperfine spectrum in Fig.~\ref{fig:MOT-scheme}c, and allow for any remaining transitions to be adequately repumped by off-resonant light.

We plot the resulting steady-state radial force experienced by molecules as a function of on-axis velocity and magnetic-field strength (which is a proxy of particle position, given the trap's magnetic-field gradient) in Fig.~\ref{fig:MOT-results}a. The magnetic-field direction is along the particle velocity, co-propagating laser frequencies have a common phase, and we average over 20 randomized phases between counter-propagating and orthogonal beams~\cite{Kogel2026}. The radial forces as a function of velocity and magnetic-field strength exhibit a structure similar to simulations of cold and strongly confining conveyor-belt MOTs in bosonic species~\cite{Yu2026,Lyu2026}. 

The forces near the MOT center are plotted in Fig.~\ref{fig:MOT-results}b, and compared with the results for \BaFOneThreeEight\ simulated with the parameters of a recent experiment realizing direct loading from a molecular beam~\cite{Zeng2026}. Considering identical intensities per frequency component, the results predict similar force magnitudes and capture velocities of up to 5\,m/s using classical trajectory simulations for both isotopologues. Together with the bichromatic slowing scheme presented above, this approach thus provides a realistic route towards direct loading of \BaFOneThreeSeven\ molecules into a conveyor-belt MOT with efficiencies comparable to those of its simpler bosonic counterpart.

\subsection{Experimental realization of the optical cycle}

So far, we have limited the discussion to the main $X(\nu=0, N=1)\rightarrow A(\nu'=0, J'=1/2^+)$ cooling transition in \BaFOneThreeSeven. However, any realistic scheme for slowing and trapping also needs to address the relevant additional electronic or vibrational decay channels through repumping lasers to minimize branching losses. 

For BCFs, spontaneous emission is relatively suppressed due to the stimulated nature of these forces and only a small number of photons are scattered. However, some molecules may still decay into ro-vibrational dark states. The optical cycle can be closed without affecting the magnitude of the forces with the use of repumping frequencies that do not couple to the states involved in the bichromatic transition. For the first vibrational repumper, for example, this can be achieved by repumping on the $X^2\Sigma^+(\nu=1) \rightarrow B^2\Sigma^{+}(\nu'=0)$ transition. 

For the MOT, where significantly more photons, on the order of $10^4$, need to be scattered, a key experimental challenge is posed by a low-lying metastable $A'{}^2\Delta$ state in BaF. This state leads to two-photon decay that breaks the rotational closure, leading to frequent decays into unwanted rotational states~\cite{Yeo2015,Collopy2018,Zeng2024}.

In odd isotopologues of BaF, as well as similar species investigated for nuclear symmetry violation tests, such as RaF, YbF, or YbOH~\cite{Kogel2026,Zeng2023}, hyperfine mixing between the $A^2\Pi_{1/2}(J=1/2^+)$ and $A^2\Pi_{1/2}(J=3/2^+)$ states leads to additional leakage in the rotational cycle, resulting in decays to similar rotational states in the electronic ground state as for the $A'{}^2\Delta$ leak channel. In \BaFOneThreeSeven, the hyperfine-induced leakage relevant for a MOT is predominantly to the $N=3$ levels in $\nu=0$ and $\nu=1$.

In Fig.~\ref{fig:MOT-scheme}b, we illustrate an efficient repumping scheme that addresses both of these leak channels. It expands on an earlier scheme developed and demonstrated with YO, and later used with \BaFOneThreeEight~\cite{Collopy2018,Deng2025,Zeng2024}. Rotational repumping up to $N=3$ in $\nu=0$ and $\nu=1$ is required to sufficiently close the optical cycle. States with $N\ge1$ can be optically repumped, while microwaves can be used to remix populations between $N=0$ and $N=1$ manifolds. 
The added challenge in odd isotopologues is that, in addition to the usual hyperfine structure of the even isotopologues, each repumping laser would now have to address the two $G$-manifolds, which are split by approximately 4.6\,GHz. Including all $m_F$ sublevels, this raises the total number of states involved in the MOT and repumping scheme to around 896. Note, however, that the number of distinct laser frequencies required remains far smaller, since each frequency component addresses many $m_F$ sublevels simultaneously and closely spaced hyperfine levels can be covered by a single broadened or modulated beam. Recent advances in optical spectrum synthesis using serrodyning and combinations of electro- and acousto-optic modulators thus render the required laser system experimentally straightforward~\cite{Rockenhaeuser2024,Kogel2025serrodynes,Holland2021}.

These results indicate that the additional hyperfine complexity of fermionic, odd isotopologues does not prevent efficient slowing or magneto-optical trapping. Furthermore, MOT temperatures of 240\,$\mu$K have been demonstrated with \BaFOneThreeEight~\cite{Zeng2026}, and gray-molasses cooling is expected to reduce this to the tens-of-$\mu$K regime, with temperatures as low as $\sim 4\,\mu$K demonstrated in comparable molecules~\cite{Burau2023,Cheuk2018}. We expect a similar temperature range for \BaFOneThreeSeven. This brings a class of molecules sensitive to nuclear symmetry-violation effects such as NSD-PV, as well as nuclear Schiff and magnetic quadrupole moments, into an experimentally accessible regime. The resulting trapped ensembles provide the starting point for the high-fidelity state preparation developed in the following section.

\begin{figure}
    \centering
        \includegraphics[width=\linewidth]{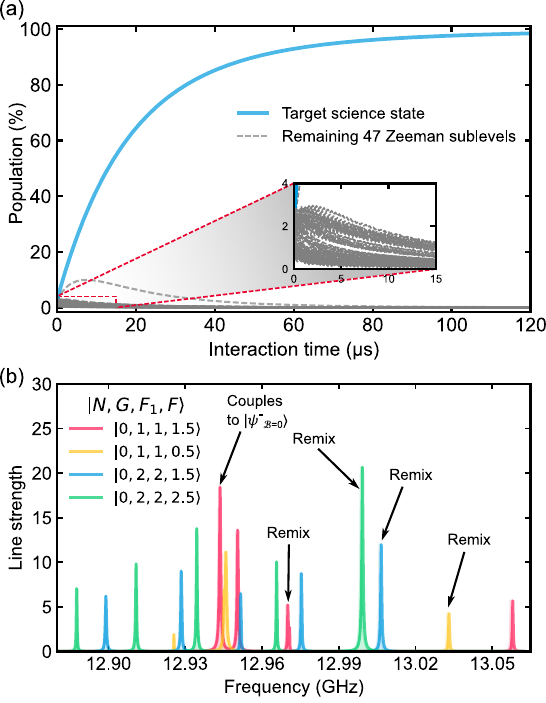}
    \caption{Science-state preparation by optical pumping. (a) Zeeman sublevel--resolved state populations for $\nu = 0, N=1$, solved with rate equations for the proposed state-preparation scheme in \BaFOneThreeSeven. The population starts from a uniform distribution across 48 Zeeman sublevels (gray dashed lines), illustrated in the zoomed inset, and is effectively pumped to the field-free science state $\ket{\psi_\mathrm{\mathcal{B}=0}^-}$ within 100\,$\mu$s (blue solid line). (b) Microwave spectrum for $N=0 \rightarrow N=1$ rotational transitions in \BaFOneThreeSeven\ showing the transitions that can be used for rotational pumping. The colors denote the four hyperfine levels in $N=0$, each of which must be addressed for complete remixing. Only one transition couples to the science state $\ket{\psi_\mathrm{\mathcal{B}=0}^-}$ in $N=1$, as annotated, and needs to be avoided when choosing a remixing scheme. The transitions labeled as \enquote{remix} can be used to address all the hyperfine levels in $N=0$ without perturbing the engineered dark science state, thereby promising effective rotational pumping when used in combination with the optical pumping in (a). This rotational pumping scheme is not included in the rate equations in (a).}
    \label{fig:state-preparation}
\end{figure}

\section{Optical pumping and state preparation}
\label{sec:state-preparation}

The cooling and trapping scheme developed above produces a large ensemble of cold molecules suitable for precision measurements, which can be further enhanced by magnetic compression and additional gray-molasses cooling~\cite{Burau2023}. However, the population remains distributed over all sublevels of $N=0$ and $N=1$ states in the electronic and vibrational ground state, which in the case of \BaFOneThreeSeven\ amount to a total of $64$ Zeeman sublevels. In contrast, typical precision sensing protocols require initialization into specific $m_F$ sublevels or their coherent superpositions. The NSD-PV measurement discussed here, in particular, requires the molecules to be prepared in either of the stretched $m_F$ states $\ket{\psi_\mathrm{\mathcal{B}=0}^+}$ or $\ket{\psi_\mathrm{\mathcal{B}=0}^-}$. 

To maximize population in these states, we develop an optical pumping scheme that initializes the population from $N=1$ to our field-free science state
$\ket{\psi_\mathrm{\mathcal{B}=0}^-}=\ket{N=1,G=1,F_1=2,F=5/2,m_F=+5/2}$ with high fidelity after the MOT. The molecules initialized in this state can subsequently be transferred adiabatically into the finite magnetic field required for the precision measurement~\cite{DeMille2008}. To further increase populations, we also discuss the possibility of combining this scheme with microwave-based rotational repumping. 

\subsection{Preparing the science state}

The central idea is to engineer the desired science state as the only dark state of the optical pumping cycle~\cite{Ho2023}. The two $G$-manifolds are addressed separately by one laser each, which we consider to be co-propagating. The $\ket{N=1, G=1,F_1,F,m_F}$ manifold is addressed using a laser with a single $\sigma^+$-polarized frequency component that is resonant with the $X^2\Sigma^+(N=1, G=1,F_1=2,F=5/2,m_F) \rightarrow A^2\Pi_{1/2}(J'=1/2^+,F_1'=2,F'=5/2,m_F')$ transition with relatively high power to optically pump into the stretched state. As a result, all other transitions in the $G=1$ manifold are driven off-resonantly by this laser. The target science state $\ket{\psi_\mathrm{\mathcal{B}=0}^-}$ is the only dark $m_F$ sublevel formed since all remaining hyperfine levels in this $G$-manifold have $F<F'_{\text{max}}=5/2$ and thus remain bright to the $\sigma^+$ light employed. 

For $\ket{N=1,G=2,F_1,F,m_F}$, rather than employing magnetic remixing to destabilize unwanted dark $m_F$ sublevels, we repump this manifold through polarization switching using a Pockels cell and multiple laser sidebands on the corresponding laser. A weak magnetic field is applied along the propagation direction of the pumping beams. This field is sufficiently weak that it only defines a quantization axis but plays no role in maintaining optical cycling. Any vibrational repumpers may also be modulated with the same Pockels cell. 

We investigate this scheme using rate equations, including only the states from $N=1$ in $\nu=0,1$ for simplicity. For the polarization-switched lasers we assume optical spectra identical to those used in our previously demonstrated optical cycling experiment~\cite{Kogel2025lasercooling}. Results are summarized in Fig.~\ref{fig:state-preparation}a. The rate-equation model predicts over 99$\%$ optical pumping into the desired stretched state within 100\,$\mu$s and with only approximately 20 scattered photons per molecule. While heating from spontaneous emission during pumping amounts to under 2\,$\mu$K increase in temperature, absorption recoils from unidirectional pumping lasers may cause significant acceleration. We find that using retro-reflected pumping beams suppresses all heating effects to the $\mu$K level, thereby preserving the temperatures reached after molasses cooling. 

\subsection{Rotational repumping}
The rate-equation model above considers only the $N=1$ manifold. Beyond this, population initially residing in $N=0$ may be transferred into the optical pumping cycle without perturbing the engineered dark state. This can be achieved using microwaves that couple the rotational manifolds, with the corresponding spectrum shown in Fig.~\ref{fig:state-preparation}b. Complete remixing requires addressing at least one transition from each of the four hyperfine levels in $N=0$, while avoiding the single transition that couples to the dark state. Suitable transitions satisfying both conditions exist for every hyperfine level. Since all these transitions have either $F'=F$ or $F'=F+1$, driving them with $\pi$-polarized microwaves additionally suppresses any undesired dark sublevels in $N=0$.

\subsection{Applicability to beam experiments, other species, and science states}

In the absence of state preparation, the statistics of a measurement such as the one proposed here are limited by the natural fraction of molecules that occupy a single Zeeman sublevel, which is only $f\approx10^{-2}$ for typical candidate species for NSD-PV measurements~\cite{DeMille2008}.

This limitation affected the proof-of-principle molecular-beam NSD-PV measurement of Ref.~\cite{Altuntas2018}, which employed the same class of science states as considered here. The optical pumping scheme developed above is therefore directly applicable and could increase the usable molecule number by up to two orders of magnitude, corresponding to roughly an order-of-magnitude improvement in statistical sensitivity. 

We note that efficient state-preparation schemes often exist only for stretched states. Identifying an efficient optical pumping scheme may be more challenging if a measurement employs states with lower or even vanishing angular momentum, for instance, to access other level crossings in an NSD-PV measurement, engineer external-field-insensitive states~\cite{Takahashi2023} or perform co-magnetometry in $\mathcal{CP}$-violation searches~\cite{Wu2020}. In such situations, combining the optical pumping scheme developed here with advances in driving high-fidelity Raman transitions between hyperfine qubits as well as the use of microwave sweeps for coherent population transfer would provide robust and broadly applicable state-preparation protocols~\cite{Holland2023,Williams2018,Ho2023}. In particular, ${}^{223}$RaF and ${}^{173}$YbF exhibit level structures similar to \BaFOneThreeSeven, and the scheme for optical pumping to a stretched state followed by coherent population transfer described here should also be applicable to these species. 

In summary, the ability to prepare nearly the entire molecular ensemble in a single Zeeman sublevel removes one of the principal statistical limitations of previous beam experiments and allows trapped-molecule measurements to fully capitalize on the increased molecule numbers achieved through laser cooling and trapping. 

\section{Dipole trapping, transport,\\measurement, and detection}
\label{sec:dipole-trapping-and-transport}

Once the molecules have been prepared in the desired science state, they must be transported into an environment that preserves coherence while enabling precise control of the electromagnetic fields required for the measurement.

For this, the ensemble can be loaded into a movable $1064$--nm optical lattice formed by two interfering laser beams~\cite{Schmid2006}. This transport scheme closely follows the optical conveyor approach demonstrated for CaF~\cite{Bao2022}. Varying the beams' relative detuning transports the molecules into the interaction region, where a stable, locally homogeneous field $\mathbf{B}$ tunes the science state-pair to near-degeneracy. The transport is adiabatic, so that the low-field eigenstates 
$\left|\psi_{\mathcal{B}=0}^{\pm}\right\rangle$ evolve smoothly into the corresponding high-field eigenstates $|\psi^{\pm}\rangle$.

As discussed in Sec.~\ref{sec:laser-cooling-and-trapping}, the expected molecular temperatures after laser cooling are in the tens-of-$\mu$K regime, which motivates a lattice depth of $U/h = 1$\,MHz, corresponding to $k_B \times 50\,\mu$K and thus several times the molecular temperature. The achievable cooling performance therefore directly determines the minimum practical trap depth and, consequently, the differential light shifts that ultimately limit the coherence time of the NSD-PV measurement, as discussed below and in Sec.~\ref{sec:sensitivities-and-limits}.

The Stark-interferometry sequence of Sec.~\ref{sec:measurement-principle} is then performed with molecules trapped in the lattice potential inside the magnetic field. A parity-selective pulse depleting $\ket{\psi^+}$ is used to initialize $\ket{\psi^-}$. Subsequently, field plates apply the oscillating $\mathcal{E}$, and a second pulse depletes $\ket{\psi^-}$ so that only the weak-interaction-populated $\ket{\psi^+}$ remains. 

In the high field, \textit{in situ} fluorescence detection of the resulting population is limited to roughly one photon per molecule~\cite{Lasner2018}. It is therefore beneficial to adiabatically transport the ensemble back to the low-field MOT region after the measurement, where a large photon cycling budget is restored, enabling high-fidelity detection. 

Following these general considerations, three practical points are crucial for successful transport and optical trapping: magnetic-field control during transport, suppression of differential light shifts, and optimization of the trapping wavelength. In the following, we discuss each in turn.

\begin{figure*}
    \centering
        \includegraphics[width=\linewidth]{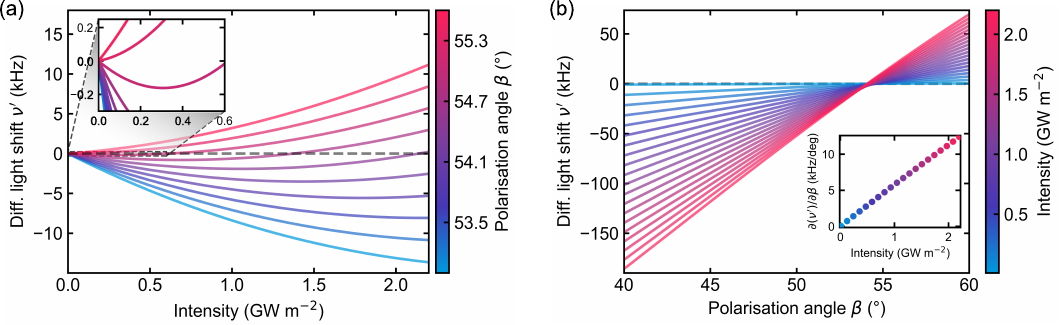}
    \caption{Magic angle trapping of NSD-PV science state-pair in \BaFOneThreeSeven. (a) Differential light shifts, $\nu'$, as a function of trapping laser intensity $I$ at $\mathcal{B}=3215$\,G for various angles $\beta$ between the laser polarization and the magnetic field. At a peak trap intensity of 2.2\,GW/m$^2$, the local extremum can be tuned to the peak intensity using $\beta$ to limit the total differential light shifts under 5\,kHz. For shallower trap depths, illustrated in the inset, the light shifts can be considerably reduced, with a peak trap intensity of 0.4\,GW/m$^2$ (corresponding to a $U/h=1$\,MHz deep trap) limiting the shifts to under 300\,Hz. (b) Differential light shifts as a function of $\beta$ for different laser intensities near their zero-crossing. The inset plots the slope of these curves as a function of laser intensity, which is used to calculate an upper bound on coherence times achievable in different lattice depths for a polarization uncertainty of $\delta \beta/\beta=10^{-4}$ in the main text.}
    \label{fig:magic-angle-trapping}
\end{figure*}

\subsection{Transport in a magnetic-field gradient}

During optical transport, two important conditions must be satisfied. 

First, the magnetic-field gradient, $\mathbf{B'}=\mathbf{\nabla |B|}$, between the MOT and the interaction region will exert a force $\mathbf{F_{B'}}/h = -g_S m_S \mu_B \mathbf{B'}/h\approx \mathbf{B'} \times 1.4$\,MHz/G on the molecules. In general, this force field can be complex due to the strong local variations of both the magnetic field direction and magnitude. The moving optical lattice must be deep enough to sufficiently counteract this force, most notably at the edge of the interaction region where $\mathcal{B}'$ can be particularly large. Approximating the effect for our $1064$--nm lattice, the depth changes by less than $10\%$ as long as $\mathcal{B}' < 2.6$\,kG/cm.

Confinement in the transverse direction is weaker, making it susceptible to even modest transverse gradients on the order of tens of G/cm. Alignment of the lattice beams along a symmetry plane of the magnetic field can suppress the transverse gradient significantly, leaving only the magnetic-field curvature. If the molecules are prepared in a high-field-seeking state, such as $\ket{\psi^-_{\mathcal{B}=0}}$ used in this work, they would be confined more strongly towards the center of curvature.

Second, since we perform state preparation before transport, the latter must remain adiabatic throughout to evolve $\ket{\psi_\mathrm{\mathcal{B}=0}^-}$ into $\ket{\psi^-}$. This requires $\dot{\mathcal{B}}/\mathcal{B} \ll \omega_{\mathrm{Larmor}} = g_S\mu_B\mathcal{B}/\hbar$, where $\dot{\mathcal{B}}=d\mathcal{B}/dt$. Magnetic fields on the order of a few Gauss already correspond to MHz-scale $\omega_{\mathrm{Larmor}}$. Thus, the weak magnetic field in the state-preparation region, together with the large fields in the regions where high gradients are expected, easily satisfy the condition required for adiabatic transport as long as the magnetic-field strength never vanishes.

\subsection{Suppression of differential light shifts}

In the interaction region, a high coherence time $\tau=1/\delta \Delta$ is desired, where $\delta \Delta$ is the temporal fluctuation and spatial inhomogeneity of the detuning between the science state-pair from degeneracy. The coherence time is thus governed by differential shifts of the molecular energy levels.

The dipole-trapping light induces an AC-Stark shift of the molecular states through scalar ($\alpha^{(0)}$), vector ($\alpha^{(1)}$), and tensor ($\alpha^{(2)}$) polarizability contributions, whose relative weights depend on the polarization of the trapping light. The scalar term is isotropic and shifts all states within an electronic manifold equally, providing the dominant contribution to the trapping potential. The vector term acts as an effective magnetic field and enters only through residual ellipticity of the nominally linear trapping polarization~\cite{Romalis1999}. The tensor term is anisotropic and depends on the $N$ and $m_N$ quantum numbers in the decoupled basis: it vanishes for $N=0$, but is finite for $N=1$, and thus constitutes an intensity-dependent differential light shift between our science states that contributes to the uncertainty $\delta \Delta$ and effectively limits $\tau$.

Clean linear polarization has previously been achieved to the $10^{-10}$ level of purity~\cite{Zhu2013}, strongly suppressing the vector light shift. Additionally, in our geometry, where the trapping laser propagates orthogonally to the quantization axis set by the magnetic field, vector light shifts are further suppressed. The resulting residual vector shift has been estimated to be in the mHz range for experimental parameters similar to those considered here, and is thus well below the dominant systematics discussed in the following~\cite{Fitch2021methods}.

The first-order tensor shifts can be suppressed by employing a certain magic angle polarization of the trapping light~\cite{Guan2021,Caldwell_Raman2020}. Figure~\ref{fig:magic-angle-trapping}a shows the differential light shift, $\nu'$, between $\ket{\psi^+}$ and $\ket{\psi^-}$ as a function of trapping laser intensity at $\mathcal{B}=3215$\,G for various angles $\beta$ between the laser polarization and the magnetic field. The calculation was performed by extending the framework developed in Ref.~\cite{Caldwell_Raman2020} to molecules with two nuclear spins (see Appendix~\ref{app:matrix-elements}), and using dynamic polarizability values for BaF at $1064\,$nm, $\alpha_{\parallel}=288$\,a.u. and $\alpha_{\perp}=671$\,a.u., calculated by \textit{ab initio} theory in Ref.~\cite{Schellenberg2026}. As an extreme example, a 5--MHz deep trap (equivalently $\sim 250$\,$\mu$K) requires a peak intensity of 2.2\,GW/m$^2$, based on which our results predict light shifts at least on the order of a few kHz. For shallower traps, as shown by the inset, the light shifts can be considerably reduced.

Using CaF molecules in optical tweezers, a rotational coherence time of up to 93\,ms has been demonstrated by tuning a local extremum in differential light shifts to the peak trapping intensity, even in the presence of 50\,Hz of differential light shifts over the entire range of the tweezers' intensity profile~\cite{Burchesky2021}. The suppression of decoherence is based on the observation that molecules only sample a fraction, $\delta I$, of the entire intensity profile of the optical trap, and thus the effective first-order differential shift $\delta \nu'_{\mathrm{I}}=\frac{\partial \nu'}{\partial I} \delta I$ could be minimized by suppressing the slope $\frac{\partial \nu'}{\partial I}$ in the intensity regime where the molecules spend most of their time. Thus, the optimal position over intensity for the local extremum depends upon the temperature of the molecular sample and its distribution within the dipole trap. Extrapolating from the results for CaF, the total differential shifts in the \BaFOneThreeSeven\ science state-pair are found to be about a factor of $6$ larger ($300\,$Hz) at a local extremum for a trap depth of $1\,$MHz ($\sim 50\,\mu$K), and would thus allow for a coherence time of up to $15$\,ms. In terms of the effective differential frequency shifts sampled by the molecules, this corresponds to $\delta \nu'_{\mathrm{I}}\approx 10$\,Hz, which feeds into the overall broadening $\delta \Delta$. Deeper traps will impose significantly stricter limits, with our estimate for a 5--MHz deep trap being limited to about a millisecond of coherence time. 

Another way for differential light shifts to contribute to $\delta \Delta$ is via polarization uncertainty $\delta \beta$ in applying the magic angle. In Fig.~\ref{fig:magic-angle-trapping}b, we calculate the differential light shifts, $\nu'$, as a function of polarization angles, $\beta$, for different intensities near their zero-crossing. We fit a linear curve through each line, and extract the light shift per degree angle, $\partial \nu'/\partial \beta$, and plot them against intensities in the inset. Its contribution to $\delta \Delta$ is given by the effective frequency shift $\delta \nu'_{\mathrm{\beta}}=\frac{\partial \nu'}{\partial \beta} \delta \beta$. Using modern Glan-type polarizers that can achieve a $\delta \beta/\beta$ of $10^{-4}$~\cite{Norrgard2019}, and given a peak intensity of 0.4\,GW/m$^2$ corresponding to a 1--MHz deep trap, we expect a $\delta \nu'_{\mathrm{\beta}} \approx 15$\,Hz and associated limit on coherence time up to 10\,ms.

\subsection{Magic wavelength trapping}
\begin{figure}[tb]
    \centering
    \includegraphics[width=1\linewidth]{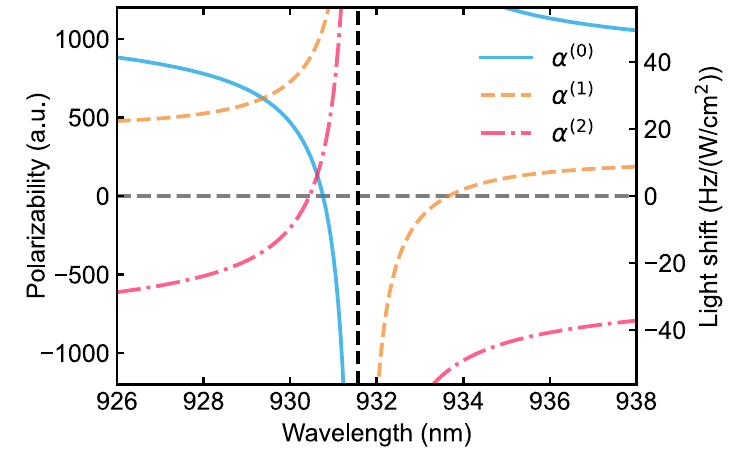}
    \caption{Calculated wavelength dependence of the scalar ($\alpha^{(0)}$), vector ($\alpha^{(1)}$) and tensor ($\alpha^{(2)}$) polarizabilities of BaF  (blue solid, orange dashed and pink dot-dashed lines, respectively) in the vicinity of the narrowline $X^2\Sigma^+ \rightarrow A'^2\Delta$ electronic transition, represented by the vertical black dashed line. Anisotropic effects dependent on rotations are not included~\cite{Guan2021}. The tensor polarizability $\alpha^{(2)}$ crosses zero at approximately 930.4\,nm, where differential tensor light shifts between a given pair of rotational states are near-perfectly suppressed.}
    \label{fig:magic-wavelength}
\end{figure}

Magic wavelength trapping for rotational qubits has been demonstrated to be a powerful alternative to magic angle approaches by tuning the molecule's tensor polarizability to zero~\cite{Kotochigova2010, Guan2021, Bause2020, Ruttley2025, Gregory2024}. In RbCs, a low-lying metastable state enabled this by offering a zero-crossing of the wavelength-dependent tensor polarizability. The crossing occurs between the vibrational poles of the near-forbidden transition, allowing for suppressed photon-scattering losses and second-scale rotational coherence times. The $A'{}^2\Delta$ state in BaF offers a similar unique situation. Assuming a natural linewidth of $\Gamma=2\pi\times 30$\,kHz and a transition dipole moment of 0.272\,a.u. from \textit{ab initio} theory~\cite{Hao2019}, we estimate the dynamic polarizabilities of BaF as a function of laser wavelength in the vicinity of the $X^2\Sigma^+ \rightarrow A'^2\Delta$ electronic transition in Fig.~\ref{fig:magic-wavelength}. The calculation was performed by adapting the toolbox developed in Ref.~\cite{Humphreys2025}. We find that the tensor polarizability crosses zero at approximately 930.4\,nm. In this region, up to some rotation-dependent anisotropic corrections, we may expect to find a wavelength where tensor light shifts between a given pair of rotational states can be near-perfectly suppressed~\cite{Guan2021}. We note that a quantitative analysis for a specific state pair requires an extended treatment including these corrections~
\cite{Guan2021}. Since the scalar polarizability is still positive, this magic wavelength would allow for high-field-seeking optical trapping. 

In a 1--MHz deep trap, from the wavelength dependence of $\alpha^{(2)}$ near the magic wavelength, we estimate an upper bound on differential light shifts between the NSD-PV science state-pair to be $\pm 1$\,Hz per 1\,MHz of trapping laser frequency drift. This allows for a second-scale coherence time for a modestly frequency-stabilized laser. Off-resonant photon scattering is then expected to be the dominant source of decoherence. Summing the off-resonant scattering contributions from the $A^2\Pi$ and
$A'^2\Delta$ states at the magic wavelength, we estimate a scattering rate of $5\,\mathrm{s}^{-1}$ for a 1--MHz deep trap, corresponding
to a scattering-limited lifetime of up to $\sim200\,\mathrm{ms}$.

Note that, as in RbCs, the vector polarizability near this magic wavelength is comparable to or larger than the scalar one (Fig.~\ref{fig:magic-wavelength}). Achieving the predicted coherence times therefore requires correspondingly good control of the trapping light polarization and the laser's alignment with the magnetic field, a regime that has already proven experimentally accessible~\cite{Ruttley2025,Gregory2024}.

These estimates suggest a promising avenue for achieving long rotational coherence times in laser-coolable molecules, and should be explored in further detail via \textit{ab initio} calculations of the molecular dynamic polarizabilities. We expect comparable magic-wavelength conditions to exist in YO molecules, which exhibit a similar $A'^2\Delta$ state and are suitable for eEDM searches and studies of many-body physics~\cite{Augenbraun2020,Mehling2025}. A similar setting also exists in YbF molecules used for eEDM searches, where 4f-hole states provide similarly narrow transitions~\cite{Popa2024}.

In summary, magic-angle trapping provides a near-term route to few-millisecond coherence times, while magic-wavelength trapping offers a potential route to second-scale coherence. As discussed above, these limits are ultimately constrained by the lattice depth, and therefore by the achievable molecular temperatures. The coherence-time limits calculated here, together with residual field inhomogeneities, feed directly into the final sensitivity limits discussed next.

\section{Systematics, sensitivities\\and limits}
\label{sec:sensitivities-and-limits}

The preceding sections have developed each component of the experimental platform individually, from the production of the cold molecular ensemble to coherent state preparation and dipole trapping. We now combine these elements to estimate the performance of the complete experimental sequence and identify the parameters that ultimately determine its sensitivity. 

\subsection{Uncertainty in $\Delta$}
The leading limit in precision for the NSD-PV measurement comes from $\delta \Delta$. The three main contributors to this quantity are the laser intensity and laser polarization uncertainties in the optical lattice, with their associated differential light shifts, $\delta \nu'_{\mathrm{I}}$ and $\delta \nu'_{\mathrm{\beta}}$, as well as the magnetic-field stability over space and time $\delta \mathcal{B}$, which leads to a shift $\delta \nu'_{\mathcal{B}}=g_S\mu_B \delta \mathcal{B}/h$. 

As motivated in Secs.~\ref{sec:laser-cooling-and-trapping} and \ref{sec:dipole-trapping-and-transport}, we adopt a lattice depth of $U/h = 1$\,MHz for the sensitivity estimates that follow.

In Sec.~\ref{sec:dipole-trapping-and-transport}, we detailed the use of magic angle polarization conditions to estimate a $\delta \nu'_{\mathrm{I}}\approx 10$\,Hz and $\delta \nu'_{\mathrm{\beta}}\approx 15$\,Hz at this depth. While the use of magic-wavelength conditions is projected to suppress these uncertainties to the sub-Hz levels, we take them into account for a conservative estimate of a near-term experiment. These trap-depth-dependent light shifts, together with the magnetic-field stability discussed below, set the achievable coherence time.

Magnetic-field control at the level of $\delta \mathcal{B}/\mathcal{B} \approx 10^{-8}$ has already been demonstrated in a beam experiment over a spatial extent of about $6\,$cm~\cite{AltuntasPRA}. The obvious benefit of a measurement in a trapped ensemble is improved field control capacity over the small size, hundreds of micrometers, of an optically trapped molecular cloud. Thus, assuming a magnetic-field homogeneity and stability of $\delta \mathcal{B} < 1$\,nT
over the $\sim(250\,\mu\mathrm{m})^3$ trap volume, corresponding to
$\delta \mathcal{B}/\mathcal{B} \approx 3 \times 10^{-9}$, within a factor of $\sim3$ of the $10^{-8}$ demonstrated over the far larger spatial extent of a beam experiment~\cite{Altuntas2018}, and over timescales $1/\omega < t < \tau$, where $\tau$ is the total duration of coherent measurement, we find
$\delta\nu'_{\mathcal{B}} \approx 28$~Hz. Although this is a stringent requirement, commercial superconducting solenoids used in precision Penning-trap experiments routinely achieve relative field stabilities at or below this level~\cite{MarieJeanne2008,Myers2013,Hanneke2011}. Complementary to the large trapped ensembles considered here, such Penning-trap systems have themselves been proposed as hosts for NSD-PV measurements with single molecular ions~\cite{Karthein2024}, trading molecule number for exceptional field stability and access to rare and short-lived nuclei. Notably, the leading systematic effects differ substantially between optical traps, Penning traps, and molecular beams~\cite{AltuntasPRA}, so that consistent results across all three complementary platforms —-- with beam measurements enhanced by the state-preparation techniques of Sec.~\ref{sec:state-preparation} --— would provide a powerful cross-validation of any observed NSD-PV signal.

We combine all three contributions to $\delta \Delta$ in quadrature as $\delta \nu'= \sqrt{(\delta \nu'_{\mathrm{I}})^2+(\delta \nu'_{\mathrm{\beta}})^2+(\delta\nu'_{\mathcal{B}})^2}$, and obtain $\delta \Delta=2\pi \times \delta \nu' \approx 2\pi \times 33$\,Hz, corresponding to a coherence time $\tau=1/\delta\Delta \approx 4.8$\,ms.

We note that the same analysis --- trap-induced differential light shifts and magnetic-field instabilities combined in quadrature to set the coherence time --- applies to other trapped-molecule symmetry tests, with the individual contributions determined by the field sensitivities of the respective science states.

\subsection{Electric-field systematics}
The science signal for the NSD-PV measurement is given by Eq.~\ref{eqn:asymmetry}. While the electric-field amplitude $\mathcal{E}$ entering this expression can be controlled with high precision using established radio-frequency techniques (see Sec.~\ref{sec:measurement-principle}), an important systematic can arise from the presence of a non-reversing electric field $\mathcal{E}_{\mathrm{nr}}$. Although static electric fields alone do not produce interference in Eq.~\ref{eqn:observable}, the motion of molecules in a conservative trap sees $\mathcal{E}_{\mathrm{nr}}$ as a time-varying field that does not change direction with $\mathcal{E}$, thus mimicking the effect of $i\hbar W$. It is possible to characterize $\mathcal{E}_{\mathrm{nr}}$ with high accuracy using Stark-interferometry methods previously used in similar precision measurements~\cite{AltuntasPRA}. In particular, measurements at different level crossings give a deterministic tuning knob for $\tilde{C}/d$, constituting a powerful method for the identification of systematics~\cite{DeMille2008, Norrgard2019}. Another approach to minimize the effect of $\mathcal{E}_{\mathrm{nr}}$ would be to confine molecules in the tighter traps of a three-dimensional optical lattice.  

A complete treatment of the systematics of Stark-interference NSD-PV measurements can be found in Refs.~\cite{Altuntas2018,AltuntasPRA}. A genuinely new source of $\mathcal{E}_\text{nr}$ in trapped experiments would be the incomplete phase reversal of the oscillating electric field, but since modern signal generators offer phase resolutions better than one part in ten thousand, we believe this to be negligible. A statistical uncertainty may arise from timing jitter between the pulse initializing $\ket{\psi^-}$, the start of the oscillating electric field, and the final projection pulse, but this should be at most $0.1$\% with typical experiment control systems operating at $100\,$MHz, and likely well below that. Averaged over $100$\,hours of data taking, this would give an uncertainty of approximately $2\,$ppm, which means that this is also negligible.

\subsection{Standard quantum limit performance}
A standard-quantum-limited measurement of NSD-PV measures the weak matrix element, $W$, with precision $\delta W = 1/(\tau\sqrt{RNT})$ where $R$ is repetition rate, $N$ is the number of molecules per shot, and $T$ is the total time period over which data is collected. Taking into account various factors such as ablation-heat load on buffer-gas cells, slowing, trapping, additional cooling and transport sequence durations, as well as data acquisition latency overheads, we expect a modest repetition rate of $R=1$\,Hz. The directly loaded conveyor-belt MOT of ${}^{138}$BaF has been estimated to have $10^4$ molecules~\cite{Zeng2026}. Though we project up to an order-of-magnitude improvement in this number from efficient bichromatic slowing and transverse cooling, dipole-trap hand-off and transport efficiencies still remain sub-optimal in molecules~\cite{Wu2021,Jorapur2024,Yu2026, Bao2022}. Thus, with the similar cooling and trapping efficiencies between different isotopologues and high-fidelity state preparation detailed in the previous sections, we estimate the number of useful \BaFOneThreeEight\ molecules in the interaction region to be $N_{138} \approx 10^4$ and \BaFOneThreeSeven\ molecules to be $N_{137} \approx 10^3$, roughly in the ratio of their natural abundances. Based on this, for a 24-hour measurement with \BaFOneThreeSeven, we calculate $\delta W/2\pi \leq 0.0036$\,Hz. 

Omitting the suppressed contribution from the fluorine nucleus, the uncertainty in $\kappa$ extracted from the barium nucleus is then $\delta \kappa_{\text{Ba}}=\delta W/|\tilde{C}_{137\text{BaF}}|W_{P , \text{Ba}}$ where $|\tilde{C}_{137\text{BaF}}|$ was calculated to be 0.44 for the science crossing and $W_{P , \text{Ba}}/2\pi = 160\,$Hz~\cite{Hao2018}. For a theoretically predicted $\kappa_{\text{Ba}}=0.07$~\cite{Flambaum1997,Haxton2002,DeMille2008}, we project a precision of $\delta \kappa_{\text{Ba}}/|\kappa_{\text{Ba}}|\leq7.2\times 10^{-4}$, an under-0.1\% measurement. This statistical precision lies well below the present theoretical uncertainty on $\kappa_{\text{Ba}}$, enabling consistency checks across the many available level crossings and
across barium isotopes, as enabled by isotopologue-selective laser cooling and trapping~\cite{Kogel2024isotope}. This ensures the measurement is ultimately limited by
systematics and nuclear-theory input rather than by counting statistics.

These estimates mark an improvement of roughly two orders of magnitude from previous beam experiments~\cite{DeMille2008}. The key contributions to the projected improvement come from two sources -- one, over an order-of-magnitude longer interrogation time enabled by trapping and better field control possibilities, and two, about two orders of magnitude higher molecule numbers, enabled by modern buffer-gas sources and the optical control techniques for molecules that were outlined here. 

\subsection{Accessing $W_{P , \text{F}}$ with \BaFOneThreeEight}
Besides its benefit as a null-probe for systematics in the measurement of $\kappa_{\text{Ba}}$~\cite{Altuntas2018}, it is interesting to consider also a complementary science measurement with \BaFOneThreeEight, which has a much simpler level structure than the odd isotopologues~\cite{Kogel2026,Rockenhaeuser2024,Rockenhaeuser2023} and is therefore experimentally more easily accessible. 

Here, with data taken over 100 hours, we project a precision of $\delta W/2\pi\leq5.5 \times 10^{-4}$\,Hz which benefits from the increased $N_{138}$. Since this species only has a fluorine nuclear spin, we extract an uncertainty in $\kappa$ from the fluorine nucleus $\delta \kappa_{\text{F}}=\delta W/|\tilde{C}_{138\text{BaF}}|W_{P , \text{F}}$. With $|\tilde{C}_{138\text{BaF}}|=0.4$ in this species, $W_{P , \text{F}}/2\pi = 0.05\,$Hz, and the theoretically calculated $\kappa_{\text{F}}=-0.08$~\cite{Flambaum1997}, we project a precision of $\delta \kappa_{\text{F}}/|\kappa_{\text{F}}|\leq0.34$, sufficient for resolving the NSD-PV effects in the fluorine nucleus.

In particular, the prospect of resolving the fluorine nuclear spin is very attractive for benchmarking nuclear structure theory calculations, which are more easily tractable for lighter nuclei~\cite{Hao2020,Gardner2026,Norrgard2019}. Further, $\kappa_{\text{Ba}}$ is expected to be dominated by the anapole contribution $\kappa_a$ of a valence neutron, which scales as $\kappa_a \propto A^{2/3}$ with nuclear mass $A$, whereas
$\kappa_{\text{F}}$ is dominated by the vector-electron--axial-nucleon term $\kappa_2$ of a valence proton, which is $A$-independent.
Measurements in both nuclei, together with comparisons across barium isotopes, can help disentangle these two contributions.

\section{Conclusion}
\label{sec:conclusion}

This work establishes a realistic and experimentally accessible roadmap toward precision measurements with spinful molecules, opening the door to a broad class of searches for nuclear symmetry-violating interactions and other manifestations of physics beyond the Standard Model.

Although fermionic, odd isotopologues of laser-coolable heavy molecules possess substantially more complex hyperfine structure than their bosonic counterparts, we find that this complexity does not fundamentally limit laser slowing, magneto-optical trapping, state preparation, or coherent trapping, all of which can significantly improve the capabilities of precision measurements exploiting these isotopologues. 

For \BaFOneThreeSeven\ specifically, the combined toolbox proposed here projects a statistical
sensitivity to the barium anapole coupling below the 0.1\% level, together with the
first access to $^{19}$F contributions in \BaFOneThreeEight.

Beyond NSD-PV, the methods developed and simulated in detail here  ---  hyperfine-resolved bichromatic slowing, conveyor-belt trapping of spinful isotopologues, dark-state engineering for state preparation, and magic trapping conditions  --- transfer directly to other spinful species, including molecules containing quadrupole- or octupole-deformed nuclei required for nuclear magnetic quadrupole and Schiff moment searches. This contributes to establishing a general experimental toolbox for laboratory studies of symmetry violation in the nuclear sector~\cite{Flambaum2014,Denis2020,Grasdijk2021}.

Many of the techniques developed here are equally applicable to NSD-PV tests using molecular beams~\cite{Altuntas2018}, where we anticipate them to lead to substantial improvements through enhanced transverse cooling, laser slowing, and state preparation. 

\section*{Note added}

During the preparation of this work, we became aware of an independent proposal for realizing a magneto-optical trap of \BaFOneThreeSeven\ molecules (Ref.~\cite{Yang2026}). While that scheme is also based on the conveyor-belt trapping mechanism, it employs a reduced set of frequency components compared to the approach presented here. Our simulations indicate that the inclusion of additional hyperfine transitions leads to substantially stronger trapping forces and larger capture velocities, suggesting significantly improved loading efficiencies, and thus higher molecule numbers, for the scheme proposed here.

\section*{Acknowledgements}
We thank Felix Kogel and Marian Rockenh\"auser for contributions in the early stages of this work. We acknowledge discussion with Anastasia Borschevsky, Eifion Prinsen and Luk\'{a}\v{s} Pa\v{s}teka on magic wavelength trapping of BaF, and with Michael Tarbutt and Jack Devlin on numerical modeling of molecular laser cooling. TG acknowledges funding support from the OeAW DOC program. Computational results in this work have been achieved using the Austrian Scientific Computing (ASC) infrastructure. This research was funded in whole or in part by the Austrian Science Fund (FWF) 10.55776/PAT8306623. It has further received funding from the European Research Council (ERC), specifically, StG NEWMAT (949431) and StG UltraMeDiQs (101219560). Views and opinions expressed are however those of the author(s) only and do not necessarily reflect those of the European Union or the European Research Council. Neither the European Union nor the granting authority can be held responsible for them. 

\bibliography{biblio}

\newpage
\onecolumngrid
\appendix

\section{Potential NSD-PV measurement level crossings in \BaFOneThreeSeven}
\label{app:level-crossings-table}

The sensitivity of each level crossing in Fig.~\ref{fig:Zeeman-crossings} to NSD-PV effects is determined by evaluating the matrix elements $\tilde{C} = \bra{\psi^-} (\mathbf{S}\times\mathbf{\hat{n}}) \cdot \frac{\mathbf{I}}{I}\ket{\psi^+}$. In Tab.~\ref{tab:science-crossings}, we identify all potential science crossings in \BaFOneThreeSeven\ and report their calculated positions $\mathcal{B}_{\mathrm{cross}}$, involved states with decoupled basis labels, NSD-PV matrix elements $\tilde{C}$ for the barium nuclear spin, and the DC-Stark dipole matrix elements $d$. The g-factors used in this calculation are $g_S=2.00197$, $g_l=-0.00593$, $g_r=-0.048 \times m_e/m_p = - 2.61\times10^{-5}$, and $g_{N_{\text{F}}}=5.258$, inferred from the measured values for \BaFOneThreeEight~\cite{Cahn2014}, where $m_e$ and $m_p$ are the electron and proton mass respectively. For the barium nuclear spin, we use $g_{N_{\text{Ba}}}=0.625$ taken from Ref.~\cite{NIST_Ba}. Zeeman-tuned rotational spectroscopy on \BaFOneThreeSeven\ may provide corrections to the g-factors used here, and consequently affect the values reported in Tab.~\ref{tab:science-crossings}. In particular, the magnetic-field positions of the crossings may be expected to shift at the level of a few Gauss.

\begin{table*}[h]
\centering
\caption{Predicted NSD-PV sensitive level crossings between science states $\ket{\psi^+}$ and $\ket{\psi^-}$ in \BaFOneThreeSeven\ for $3000$--$6000$\,G, grouped by the lab-frame projection of the total angular momentum $m_F$.
State labels give the dominant decoupled basis component $(N, m_N, m_S, m_{I_1}, m_{I_2})$,
with the admixture weight in parentheses. $|\tilde{C}|$ is the magnitude of the NSD-PV matrix element for the barium nuclear spin $I_{\mathrm{Ba}}=3/2$ and
$|d|/h$ is the magnitude of the DC Stark matrix element in units of frequency shift per unit electric field.}
\label{tab:science-crossings}
\begin{ruledtabular}
\begin{tabular}{c l l c c}
$\mathcal{B}_\mathrm{cross}$ (G) & $\ket{\psi^+}$ $(N, m_N, m_S, m_{I_1}, m_{I_2})$ & $\ket{\psi^-}$ $(N, m_N, m_S, m_{I_1}, m_{I_2})$ & $|\tilde{C}|$ & $|d|/h$ (kHz/(V/cm)) \\
\hline
\multicolumn{5}{l}{$m_F = 5/2$} \\
\hline
3215.7 & $(0,0,+1/2,+3/2,+1/2)$ \;(100.0\%) & $(1,+1,-1/2,+3/2,+1/2)$ \;(97.1\%) & $0.442$ & 1.742 \\
\hline
\multicolumn{5}{l}{$m_F = 3/2$} \\
\hline
3521.3 & $(0,0,+1/2,+1/2,+1/2)$ \;(97.5\%)  & $(1,0,-1/2,+3/2,+1/2)$ \;(97.5\%) & $0.330$ & 1.915 \\
3540.8 & $(0,0,+1/2,+1/2,+1/2)$ \;(97.5\%)  & $(1,+1,-1/2,+3/2,-1/2)$ \;(97.5\%) & $0.002$ & 0.010 \\
3888.6 & $(0,0,+1/2,+1/2,+1/2)$ \;(97.8\%)  & $(1,+1,-1/2,+1/2,+1/2)$ \;(96.0\%) & $0.226$  & 3.488 \\
3220.8 & $(0,0,+1/2,+3/2,-1/2)$ \;(100.0\%) & $(1,0,-1/2,+3/2,+1/2)$ \;(97.1\%) & $0.002$  & 0.092 \\
3239.1 & $(0,0,+1/2,+3/2,-1/2)$ \;(100.0\%) & $(1,+1,-1/2,+3/2,-1/2)$ \;(97.1\%) & $0.442$ & 1.846 \\
3573.3 & $(0,0,+1/2,+3/2,-1/2)$ \;(100.0\%) & $(1,+1,-1/2,+1/2,+1/2)$ \;(95.4\%) & $0.000$ & 0.017 \\
\hline
\multicolumn{5}{l}{$m_F = 1/2$} \\
\hline
3909.4 & $(0,0,+1/2,-1/2,+1/2)$ \;(96.0\%) & $(1,0,-1/2,+3/2,-1/2)$ \;(97.8\%) & $0.002$  & 0.012 \\
3922.8 & $(0,0,+1/2,-1/2,+1/2)$ \;(96.0\%) & $(1,-1,-1/2,+3/2,+1/2)$ \;(97.6\%) & $0.096$  & 0.988 \\
4287.6 & $(0,0,+1/2,-1/2,+1/2)$ \;(96.6\%) & $(1,+1,-1/2,+1/2,-1/2)$ \;(95.9\%) & $0.034$ & 0.127 \\
4288.9 & $(0,0,+1/2,-1/2,+1/2)$ \;(96.6\%) & $(1,-1,+1/2,+1/2,+1/2)$ \;(85.2\%) & $0.381$ & 1.161 \\
4726.6 & $(0,0,+1/2,-1/2,+1/2)$ \;(97.2\%) & $(1,0,+1/2,-1/2,+1/2)$ \;(85.9\%) & $0.046$ & 4.575 \\
3545.1 & $(0,0,+1/2,+1/2,-1/2)$ \;(97.5\%) & $(1,0,-1/2,+3/2,-1/2)$ \;(97.4\%) & $0.330$ & 1.887 \\
3558.4 & $(0,0,+1/2,+1/2,-1/2)$ \;(97.5\%) & $(1,-1,-1/2,+3/2,+1/2)$ \;(97.2\%) & $0.007$ & 0.010 \\
3911.9 & $(0,0,+1/2,+1/2,-1/2)$ \;(97.8\%) & $(1,+1,-1/2,+1/2,-1/2)$ \;(95.8\%) & $0.226$ & 3.590 \\
3914.1 & $(0,0,+1/2,+1/2,-1/2)$ \;(97.8\%) & $(1,-1,+1/2,+1/2,+1/2)$ \;(85.1\%) & $0.010$ & 0.103 \\
4347.0 & $(0,0,+1/2,+1/2,-1/2)$ \;(98.2\%) & $(1,0,+1/2,-1/2,+1/2)$ \;(85.3\%) & $0.000$ & 0.019 \\
\hline
\multicolumn{5}{l}{$m_F = -1/2$} \\
\hline
4394.2 & $(0,0,+1/2,-3/2,+1/2)$ \;(96.4\%) & $(1,+1,+1/2,-3/2,-1/2)$ \;(26.5\%) & $0.000$ & 0.003 \\
4755.0 & $(0,0,+1/2,-3/2,+1/2)$ \;(96.9\%) & $(1,-1,-1/2,+1/2,+1/2)$ \;(97.1\%) & $0.099$ & 0.847 \\
4764.7 & $(0,0,+1/2,-3/2,+1/2)$ \;(97.0\%) & $(1,-1,+1/2,+1/2,-1/2)$ \;(41.9\%) & $0.002$ & 0.031 \\
5200.0 & $(0,0,+1/2,-3/2,+1/2)$ \;(97.5\%) & $(1,0,+1/2,-1/2,-1/2)$ \;(39.4\%) & $0.012$ & 0.028 \\
5203.3 & $(0,0,+1/2,-3/2,+1/2)$ \;(97.5\%) & $(1,0,-1/2,-1/2,+1/2)$ \;(97.4\%) & $0.328$  & 0.205 \\
5729.4 & $(0,0,+1/2,-3/2,+1/2)$ \;(98.0\%) & $(1,+1,-1/2,-3/2,+1/2)$ \;(100.0\%) & $0.374$  & 4.868 \\
3946.1 & $(0,0,+1/2,-1/2,-1/2)$ \;(96.0\%) & $(1,+1,+1/2,-3/2,-1/2)$ \;(26.7\%) & $0.096$  & 0.985 \\
4303.1 & $(0,0,+1/2,-1/2,-1/2)$ \;(96.6\%) & $(1,-1,-1/2,+1/2,+1/2)$ \;(96.5\%) & $0.010$ & 0.051 \\
4312.7 & $(0,0,+1/2,-1/2,-1/2)$ \;(96.6\%) & $(1,-1,+1/2,+1/2,-1/2)$ \;(41.1\%) & $0.383$ & 1.130 \\
4750.0 & $(0,0,+1/2,-1/2,-1/2)$ \;(97.2\%) & $(1,0,+1/2,-1/2,-1/2)$ \;(38.3\%) & $0.046$ & 4.680 \\
4754.2 & $(0,0,+1/2,-1/2,-1/2)$ \;(97.2\%) & $(1,0,-1/2,-1/2,+1/2)$ \;(96.9\%) & $0.001$ & 0.073 \\
5293.6 & $(0,0,+1/2,-1/2,-1/2)$ \;(97.7\%) & $(1,+1,-1/2,-3/2,+1/2)$ \;(100.0\%) & $0.000$  & 0.014 \\
\hline
\multicolumn{5}{l}{$m_F = -3/2$} \\
\hline
4778.3 & $(0,0,-1/2,-1/2,-1/2)$ \;(88.4\%) & $(1,0,+1/2,-3/2,-1/2)$ \;(37.1\%) & $0.099$ & 0.843 \\
5213.2 & $(0,0,-1/2,-1/2,-1/2)$ \;(87.3\%) & $(1,-1,-1/2,-1/2,+1/2)$ \;(97.5\%) & $0.006$ & 0.015 \\
5227.2 & $(0,0,-1/2,-1/2,-1/2)$ \;(87.3\%) & $(1,+1,-1/2,-3/2,-1/2)$ \;(43.7\%) & $0.328$ & 0.168 \\
5743.7 & $(0,0,-1/2,-1/2,-1/2)$ \;(86.2\%) & $(1,0,-1/2,-3/2,+1/2)$ \;(100.0\%) & $0.005$ & 0.120 \\
5752.8 & $(0,0,-1/2,-1/2,-1/2)$ \;(86.2\%) & $(1,0,-1/2,-1/2,-1/2)$ \;(54.2\%) & $0.374$ & 4.976 \\
\end{tabular}
\end{ruledtabular}
\end{table*}

\section{Matrix elements of relevant effective Hamiltonians}
\label{app:matrix-elements}

In this section, we provide the Hund's case $(a_{\beta S})$ matrix elements for various Hamiltonian terms that were used in the calculations reported in this work. This Hund's case is convenient for describing laser-coolable diatomic molecules with two nuclear spins, and a basis transformation allows us to extract case $(b_{\beta S})$ labels that describe the external field-free energy eigenstates of the ${}^2\Sigma$ electronic ground state, as described in Sec.~\ref{sec:measurement-principle}. In case $(a_{\beta S})$, the orbital and electron spin angular momenta, $\mathbf{L}$ and $\mathbf{S}$, are strongly coupled to form $\mathbf{\Omega}=\mathbf{\Lambda}+\mathbf{\Sigma}$ via their projections onto the internuclear axis, $\mathbf{\Lambda}$ and $\mathbf{\Sigma}$ respectively. $\mathbf{\Omega}$ couples with the rotations $\mathbf{R}$ to form $\mathbf{J}=\mathbf{\Omega}+\mathbf{R}$. The nuclear spins $\mathbf{I_1}$ and $\mathbf{I_2}$ successively couple to $\mathbf{J}$ to form the intermediate angular momentum $\mathbf{F_1}=\mathbf{J}+\mathbf{I_1}$ and the total angular momentum $\mathbf{F}=\mathbf{F_1}+\mathbf{I_2}$.

For brevity, we use $\ket{i}$ as shorthand for case $(a_{\beta S})$ states given by the quantum numbers $\ket{\eta,\Lambda,S,\Sigma,J,\Omega,F_1,F,m_F}$ associated to the angular momenta described above. Here, $m_F$ is the laboratory-frame projection of $\mathbf{F}$ and $\eta$ represents all residual quantum numbers. 

First, we consider the effective Zeeman Hamiltonian for a ${}^2\Sigma$ state with two nuclear spins given in Eq.~\ref{eqn:Zeeman-Hamiltonian}. The matrix elements are given by
\begin{equation}
    \begin{split}
        \bra{i'}H_Z\ket{i} &=
        \mu_B \mathcal{B}\,\delta_{m_F,m_F'}\sum_{q=0,\pm1}
        (-1)^{J+I_1+F_1'+1}\,(-1)^{F'+F_1+I_2+1}\,(-1)^{F-m_F}\,(-1)^{J-\Omega} \\
        &\quad\times\,[J][J'][F_1][F_1'][F][F']
        \begin{Bmatrix}J&F_1&I_1\\F_1'&J'&1\end{Bmatrix}
        \begin{Bmatrix}F_1&F&I_2\\F'&F_1'&1\end{Bmatrix}
        \begin{pmatrix}F&1&F'\\-m_F&0&m_F\end{pmatrix}
        \begin{pmatrix}J&1&J'\\-\Omega&q&\Omega'\end{pmatrix} \\
        &\quad\times\left\{(g_S+g_r+g_l)(-1)^{S-\Omega}\sqrt{S(S{+}1)(2S{+}1)}
        \begin{pmatrix}S&1&S\\-\Omega&q&\Omega'\end{pmatrix}
        -\,g_l\,\Omega\,\delta_{\Omega,\Omega'}\right\} \\[1.2em]
        &\quad-\,g_r\mu_B \mathcal{B}\,\delta_{J,J'}\,\delta_{\Omega,\Omega'}\,\delta_{m_F,m_F'} \\
        &\quad\times(-1)^{J+I_1+F_1'+1}\,(-1)^{F'+F_1+I_2+1}\,(-1)^{F-m_F}
        \,[F_1][F_1'][F][F']
        \begin{Bmatrix}J&F_1&I_1\\F_1'&J&1\end{Bmatrix}
        \begin{Bmatrix}F_1&F&I_2\\F'&F_1'&1\end{Bmatrix} \\
        &\quad\times\sqrt{J(J{+}1)(2J{+}1)}\;
        \begin{pmatrix}F&1&F'\\-m_F&0&m_F\end{pmatrix} \\[1.2em]
        &\quad-\,g_{N_1}\mu_N \mathcal{B}\,\delta_{J,J'}\,\delta_{\Omega,\Omega'}\,\delta_{m_F,m_F'} \\
        &\quad\times(-1)^{J+I_1+F_1+1}\,(-1)^{F'+F_1+I_2+1}\,(-1)^{F-m_F}
        \,[F_1][F_1'][F][F']
        \begin{Bmatrix}I_1&F_1&J\\F_1'&I_1&1\end{Bmatrix}
        \begin{Bmatrix}F_1&F&I_2\\F'&F_1'&1\end{Bmatrix} \\
        &\quad\times\sqrt{I_1(I_1{+}1)(2I_1{+}1)}\;
        \begin{pmatrix}F&1&F'\\-m_F&0&m_F\end{pmatrix} \\[1.2em]
        &\quad-\,g_{N_2}\mu_N \mathcal{B}\,\delta_{J,J'}\,\delta_{F_1,F_1'}\,\delta_{\Omega,\Omega'}\,\delta_{m_F,m_F'} \\
        &\quad\times(-1)^{F+F_1+I_2+1}\,(-1)^{F-m_F}
        \;[F][F']
        \begin{Bmatrix}I_2&F&F_1\\F'&I_2&1\end{Bmatrix}
        \sqrt{I_2(I_2{+}1)(2I_2{+}1)}\;
        \begin{pmatrix}F&1&F'\\-m_F&0&m_F\end{pmatrix} \\[1.2em]
    \end{split}
\end{equation}
where square brackets are shorthand for $[x]=\sqrt{2x+1}$, $\Lambda = 0$, and therefore, $\Omega=\Sigma$. We set $I_1 = I_{\text{Ba}}$ and $I_2 = I_{\text{F}}$, and correspondingly for the nuclear g-factors. Anisotropic and $\Lambda$-dependent terms that do not contribute in a $\Sigma$ state, but become relevant for $\Lambda \neq 0$ states, are omitted. Here and in the following, the Kronecker deltas make explicit the selection rules already contained in the 3j and 6j symbols, given by parenthesis and curly brackets, respectively, for clarity.

The DC-Stark matrix element $d$ between each science crossing governs the Stark mixing in the measurement scheme described in Sec.~\ref{sec:measurement-principle}. It is calculated from the DC-Stark Hamiltonian $H_{\mathrm{dc}}=\mathbf{D}\cdot\mathbf{E}$, where $\mathbf{E}=\mathcal{E}\hat{\mathbf{z}}$ is an external laboratory-frame electric field and $\mathbf{D}$ is the permanent electric dipole moment of the molecule along the internuclear axis. For a $\Sigma$ state, the matrix element is given by
\begin{equation}
    \begin{split}
        \bra{i'}H_\mathrm{dc}/\mathcal{E}\ket{i} &=
        D\,\delta_{\Omega,\Omega'}\delta_{J,J'\pm1,0}\delta_{F,F'\pm1,0}\delta_{m_F,m_F'\pm1,0} \\
        &\quad\times (-1)^{F+F'+F_1+F_1'+I_1+I_2+\Omega-m_F} \\
        &\quad\times[F][F'][F_1][F_1'][J][J'] \\
        &\quad\times
        \begin{Bmatrix}F_1&F&I_2\\F'&F_1'&1\end{Bmatrix}
        \begin{Bmatrix}J&F_1&I_1\\F_1'&J'&1\end{Bmatrix} \\
        &\quad\times
        \begin{pmatrix}F&1&F'\\-m_F&m_F-m_F'&m_F'\end{pmatrix}
        \begin{pmatrix}J&1&J'\\-\Omega&0&\Omega'\end{pmatrix}
    \end{split}
\end{equation}
where $D=3.17$\,Debye for BaF~\cite{Ernst1986}. 

The calculations for AC-Stark-induced differential light shifts in Sec.~\ref{sec:dipole-trapping-and-transport} were performed by extending the framework developed in Ref.~\cite{Caldwell_Raman2020} to molecules with two nuclear spins in Hund's case $(a_{\beta S})$. Here, while the prescription remains identical, the matrix elements of the molecular polarizability tensor $\mathcal{A}^K_P$ for a ${}^2\Sigma$ state are given by 
\begin{equation}
    \begin{split}
        \bra{i'}\mathcal{A}^K_P\ket{i} &= \delta_{\Omega,\Omega'}
        (-1)^{F'-m_F'+F'+F_1+I_2+K+J'+I_1+F_1+K-\Omega+J'} \\
        &\quad\times[F][F'][F_1][F_1'][J][J'] \\
        &\quad\times
        \begin{Bmatrix}F_1&F&I_2\\F'&F_1'&K\end{Bmatrix}
        \begin{Bmatrix}J'&F_1'&I_1\\F_1&J&K\end{Bmatrix} \\
        &\quad\times
        \begin{pmatrix}F'&K&F\\-m_F'&P&m_F\end{pmatrix}
        \begin{pmatrix}J'&K&J\\-\Omega&0&\Omega\end{pmatrix} \alpha^{(K)}
    \end{split}
\end{equation}
where $K$ labels rank, $P$ labels the laboratory-frame component of the laser polarization, and $\alpha^{(K)}$ is the $K$-th order polarizability.

Finally, the NSD-PV sensitivities $\tilde{C}$ are evaluated using the matrix element from Ref.~\cite{RahmlowPhD} for a ${}^2\Sigma$ state given by
\begin{equation}
    \begin{split}
        \bra{i'}\left( \mathbf{S}\times\mathbf{\hat{n}} \right) \cdot \frac{\mathbf{I_1}}{I_1}\ket{i} &= \delta_{\Omega,\Omega'\pm1} \, \delta_{J,J'\pm1,0}\,
        \delta_{F_1,F_1'}\,\delta_{F,F'}\,\delta_{m_F,m_F'} \\
        &\quad\times \frac{i\sqrt{2}\,\Omega'\,(\Omega-\Omega')}{I_1}
        (-1)^{F_1'+J+J'+I_1-\Omega} \\
        &\quad\times [J][J'] \sqrt{I_1(I_1+1)(2I_1+1)} \\
        &\quad\times
        \begin{Bmatrix}F_1'&I_1&J\\1&J'&I_1\end{Bmatrix}
        \begin{pmatrix}J&1&J'\\-\Omega&\Omega-\Omega'&\Omega'\end{pmatrix}.
    \end{split}
\end{equation}

\end{document}